\documentclass{aa}
\usepackage{txfonts}
\usepackage{graphicx,amssymb,subfigure}
\usepackage[normalem]{ulem}

\begin{document}

%
%
\newcommand{\wave}[1]{$\lambda#1\,\mathrm{cm}$}  
\newcommand{\wwave}[2]{$\lambda\lambda#1,#2\,\mathrm{cm}$}  
\newcommand{\wwwave}[3]{$\lambda\lambda#1,#2,#3\,\mathrm{cm}$}  
\newcommand{\dif}{\,\mathrm{d}}
\newcommand{\HI}{\mathrm{H\,\scriptstyle I}}
\newcommand{\HII}{\mathrm{H\,\scriptstyle II}}
\newcommand{\Brpe}{B_{\perp}}
\newcommand{\Brpa}{B_{\parallel}}
%
%
\def\cm{\,{\rm cm}\,}
\def\mum{\,\mu{\rm m}\,}
\def\ccm{\,{\rm cm^{-3}}\,}
\def\erg{\,{\rm erg}\,}
\def\kms{\,{\rm km\,s^{-1}}\,}
\def\s{\,{\rm s}\,}
\def\yr{\,{\rm yr}\,}
\def\degr{\hbox{$^\circ$}\,}
\def\farcm{\hbox{$.\mkern-4mu^\prime$}\,}
\def\farcs{\hbox{$.\!\!^{\prime\prime}$}\,}
\def\Jy{\,{\rm Jy}\,}
\def\Jyb{\,{\rm Jy/beam\,area}\,}
\def\mJy{\,{\rm mJy}\,}
\def\muJy{\,\mu{\rm Jy}\,}
\def\mJyb{\,{\rm mJy/beam\,area}\,}
\def\muJyb{\,\mu{\rm Jy/beam\,area}\,}
\def\K{\,{\rm K}\,}
\def\muG{\,\mu{\rm G}\,}
\def\MHz{\,{\rm MHz}\,}
\def\GHz{\,{\rm GHz}\,}
\def\pc{\,{\rm pc}}
\def\kpc{\,{\rm kpc}}
\def\Mpc{\,{\rm Mpc}}
\def\n{\,n_\rm{e}\,}
\def\radm{\,\mathrm{rad\,m^{-2}}\,}

\title{Magnetism in the spiral galaxy NGC~6946: magnetic arms, depolarization
rings, dynamo modes and helical fields}

\titlerunning{Magnetic arms in NGC~6946}
\authorrunning{R.~Beck}

\author{Rainer~Beck \inst{}}

\institute{Max-Planck-Institut f\"ur Radioastronomie, Auf dem H\"ugel 69,
53121 Bonn, Germany, rbeck@mpifr-bonn.mpg.de}

\offprints{R.~Beck}

\date{Received 20 December 2006 / Accepted 8 May 2007}

\abstract{The spiral galaxy NGC~6946 hosts \emph{magnetic spiral
arms}, highly aligned magnetic fields between the gas/optical arms.}
{The origin of the magnetic phenomena and their relation to the
interstellar gas are investigated.}
{NGC~6946 was observed in total intensity and linear polarization
in five radio bands between 3~cm and 21~cm. Maps of spectral index,
Faraday rotation and depolarization were derived.}
{At the inner edge of the inner gas spiral arm the ordered magnetic
field is only mildly compressed and turns smoothly, to become aligned
along the gas arm. Hence the field is not shocked and is probably
connected to the warm, diffuse gas. At larger radii,
two bright magnetic arms between the optical arms are visible in
polarized intensity. The field in the northern magnetic arm is almost
totally aligned. Faraday rotation measures ($RM$) in
these arms are consistent with the superposition of two low
azimuthal dynamo modes. Three more magnetic arms are
discovered in the outer galaxy, located between $\HI$ arms.
The $RM$ structure function confirms large-scale coherent fields.
The observed anti-correlation between the field's pitch angles and the $RM$
values is a possible signature of helical fields. --
Due to strong Faraday depolarization the galaxy is not
transparent to polarized waves at \wave{18} and \wave{20}. The
large-scale asymmetry in depolarization with respect to the major
axis may be another indication of large-scale helical fields.
Three depolarization rings of almost zero polarization seen at \wave{20}
are probably generated by differential Faraday rotation in $\HII$
complexes in NGC~6946 of 300--500$\pc$ size. --
In the gas/optical spiral arms, the total (mostly turbulent) magnetic
field is amplified to $\simeq 20~\mu$G.  Its energy density is $\simeq$10
times larger than that of the ionized gas and is similar to that of the
turbulent gas motions in the inner galaxy. The magnetic energy exceeds
that of the turbulent energy in the outer galaxy. All energy densities
in NGC~6946 are about one order of magnitude larger than those in the
Milky Way.}
{Density waves in the inner gaseous spiral arms mildly compress the field.
Dynamo action probably generates the magnetic spiral arms.
The magnetic field is dynamically important, interacts
with the gas flow and possibly determines the properties of the gas
spiral arms.}

\keywords{Galaxies: spiral -- galaxies: magnetic fields -- galaxies: ISM --
        galaxies: individual: NGC~6946 -- radio continuum: galaxies --
        radio continuum: ISM}

\maketitle


\section{Introduction}
\label{sect:intro}

Radio polarization observations revealed magnetic fields of
considerable strength in all galaxies containing a significant
amount of cold gas (see Beck\ \cite{B05} for a review). Magnetic
fields are probably important for star formation (Mac Low \&
Klessen\ \cite{M04}; Heiles \& Crutcher\ \cite{H05}), especially the
first generation of stars (Silk \& Langer\ \cite{SL06}), and for the
formation of spiral arms (Kim \& Ostriker\ \cite{K02}; G\'omez \&
Cox\ \cite{G02}), but their role is not yet understood.

NGC~6946 is a nearby grand-design spiral galaxy of Hubble type Scd
at only $\approx 5.5$\Mpc\ distance (Tully\ \cite{T88}), so that
1\arcmin\ corresponds to 1.6\kpc. With an optical diameter of about
20\arcmin\ it is one of the largest galaxies in the sky. The spiral
arms are massive, but without prominent dust lanes at their inner
edges, indicating relatively weak density waves. No companion galaxy
is known, and the small inner molecular bar can hardly excite
density waves. The distribution of gas and star formation is
asymmetric (``lopsided''), with one prominent spiral arm in the
north-east.

NGC~6946 is bright in all spectral ranges and hence an ideal
laboratory to test the interaction between the components of the
interstellar medium. The radio-infrared correlation within galaxies
was first studied there (Beck \& Golla \cite{BG88}; Bicay et al.\
\cite{BH89}). Linearly polarized radio continuum emission was first
detected at \wave{2.8} with the Effelsberg telescope (Klein et al.\
\cite{K82}), followed by more sensitive Effelsberg observations at
\wave{6.2} and \wave{2.8} (Harnett et al.\ \cite{H89}; Ehle \& Beck\
\cite{EB93}) and observations with the VLA at \wwave{18}{20} (Beck\
\cite{B91}) and \wave{6} (Beck \& Hoernes\ \cite{BH96}). The latter
authors combined VLA and Effelsberg data and demonstrated that the
missing small baselines of an interferometer reduce the large-scale
emission not only in the Stokes parameter $I$, but also in $Q$ and
$U$. Thus the combination of high-frequency data from synthesis and
single-dish telescopes is also essential for polarization
measurements.

The combined \wave{6.2} polarization map by Beck \& Hoernes
(\cite{BH96}) revealed two surprisingly symmetric ``magnetic spiral
arms'' located between the optical spiral arms. After subtraction of
the diffuse unpolarized radio background the degree of polarization
is 30--60\%. This means that the field in the magnetic arms is
almost perfectly aligned, parallel to the adjacent spiral arms. The
symmetry and alignment of the magnetic arms indicated that a new
kind of physical phenomenon was discovered, with possibly major
consequences for our understanding of galactic magnetic fields.
Analysis of the magnetic arms in NGC~6946 with 1-D wavelet functions
by Frick et al. (\cite{F00}) indicated that they are phase-shifted
``images'' of the optical arms. The cross-correlation technique
using 2-D isotropic wavelets demonstrated an anticorrelation between
polarized intensity and gas density (Frick et al.\ \cite{F01}).

After the discovery of the magnetic arms in NGC~6946, several
authors presented dynamo models with generation of magnetic arms
(Moss\ \cite{M98}; Shukurov\ \cite{S98}; Rohde et al.\ \cite{R99}).
Dynamo action is supported by the observational result that Faraday
rotation measures in NGC~6946 are preferentially positive in the
northern magnetic arm and negative in the southern arm, thus a
large-scale field with a coherent direction exists (Ehle \& Beck\
\cite{EB93}). As the northern side is nearer to us, the radial
component of the spiral magnetic field points inwards, i.e. towards
the galaxy's centre. A similar result was found for spiral fields in
several other galaxies, but neither dynamo action nor large-scale
primordial seed fields are able to explain this preference (Krause
\& Beck\ \cite{KB98}). An alternative interpretation of the magnetic
arms as slow, phase-shifted MHD waves (Fan \& Lou\ \cite{F96},
\cite{F97}; Lou \& Fan\ \cite{L98}) was intriguing. However, slow
MHD waves can exist only in the rigidly rotating part of a galaxy
while the magnetic arms are mainly observed in the outer regions.
Fundamental criticism was raised by Shukurov (\cite{S05}) who
claimed that slow MHD waves are unstable in three dimensions.

In spite of all efforts the phenomenon of magnetic arms is far from
being understood. In this paper new multi-wavelength radio
observations of NGC~6946 which were obtained during the last years
are presented and discussed. The two main magnetic arms are
confirmed at all wavelengths, but more magnetic arms exist. The
almost perfect alignment of the magnetic field vectors along the arm
structure suggests close interaction between the field and
the gas flow.

\section{Observations and data reduction}
\label{sect:obs}

\begin{table*}           
\caption{\label{Table1}Radio continuum observations of NGC~6946}
\centering
\begin{tabular}{lcc}
\hline\hline
\  & VLA & Effelsberg \\
\hline
Frequency (GHz)       & 8.435 \& 8.485  & 10.55 \\
Wavelength (cm)       & 3.55  \& 3.52   & 2.84 \\
Configuration            & D              \\
Pointings                & 3 \\
Observing dates \hfill   & 1995 Apr 13+20, 1996 July 21+23  & 1990 June to 1991 Feb \\
Net observing time (h)   & 26              & 60 \\
Resolution of final maps & 8\farcs5 \& 15\arcsec & 68\arcsec \\
Rms noise in $I$; $Q$ and $U$ ($\mu{\rm Jy/beam\,area}$)  & 10; 7 \& 15; 8  & 450; 200 \\
Reference                &                 & Ehle \& Beck (\cite{EB93}) \\
\hline

Frequency (GHz)       & 4.835 \& 4.885  & 4.85 \\
Wavelength (cm)       & 6.20  \& 6.14   & 6.18 \\
Configuration            & D              \\
Pointings                & 1 \\
Observing dates \hfill   & 1991 Mar 23 + Apr 1 & 1997 Jan to 1998 Sep \\
Net observing time (h)   & 10              & 18 \\
Resolution of final maps & 15\arcsec       & 147\arcsec \\
Rms noise in $I$; $Q$ and $U$ ($\mu{\rm Jy/beam\,area}$) & 22; 15 & 500; 70 \\
Reference                & Beck \& Hoernes (\cite{BH96}) \\
\hline

Frequency (GHz)       &                 & 2.675 \\
Wavelength (cm)       &                 & 11.21 \\
Observing dates \hfill   &                 & 1999 Apr to July \\
Net observing time (h)   &                 & 7 \\
Resolution of final maps &                 & 300\arcsec \\
Rms noise in $I$; $Q$ and $U$ ($\mu{\rm Jy/beam\,area}$)  &       & 1000; 600 \\
\hline

Frequency (GHz)       & 1.465 \& 1.665  & 1.400 \\
Wavelength (cm)       & 20.46 \& 18.01  & 21.41 \\
Configuration            & D + C          \\
Pointings                & 1 \\
Observing date  \hfill   & 1988 July 26 + 1996 Apr 4  & 1998 Sep \\
Net observing time (h)   & 12 +  11       & 7 \\
Resolution of final maps & 15\arcsec       & 560\arcsec \\
Rms noise in $I$; $Q$ and $U$ ($\mu{\rm Jy/beam\,area}$) & 20; 12 & 5000; 2000 \\
Reference                & Beck (\cite{B91}) (D array) \\
\hline

\label{tab1}
\end{tabular}
\end{table*}

\begin{figure*}[htbp]
\includegraphics[bb = 62 157 528 616,width=0.465\textwidth,clip=]{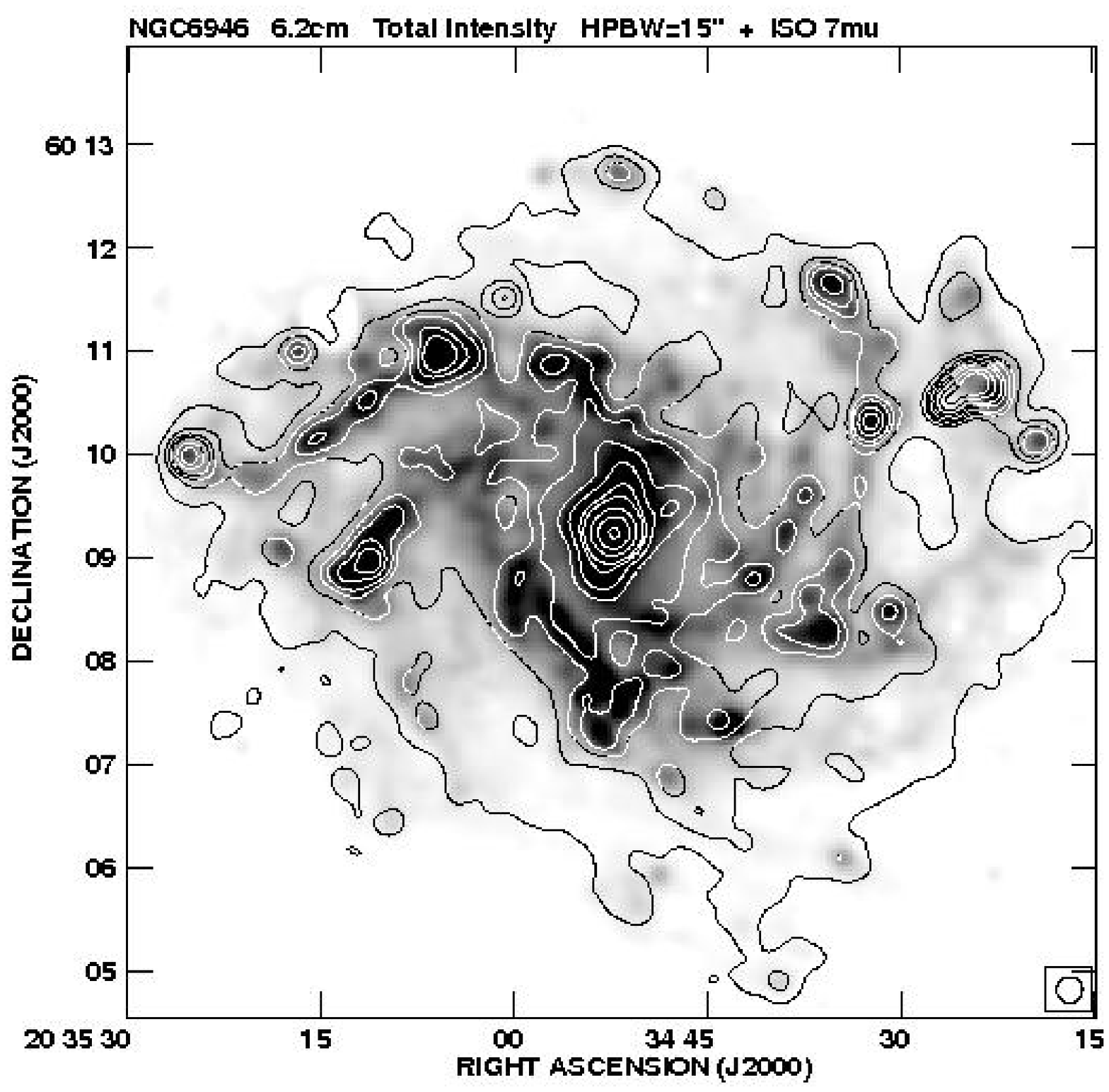}
\hfill
\includegraphics[bb = 62 128 528 623,width=0.465\textwidth,clip=]{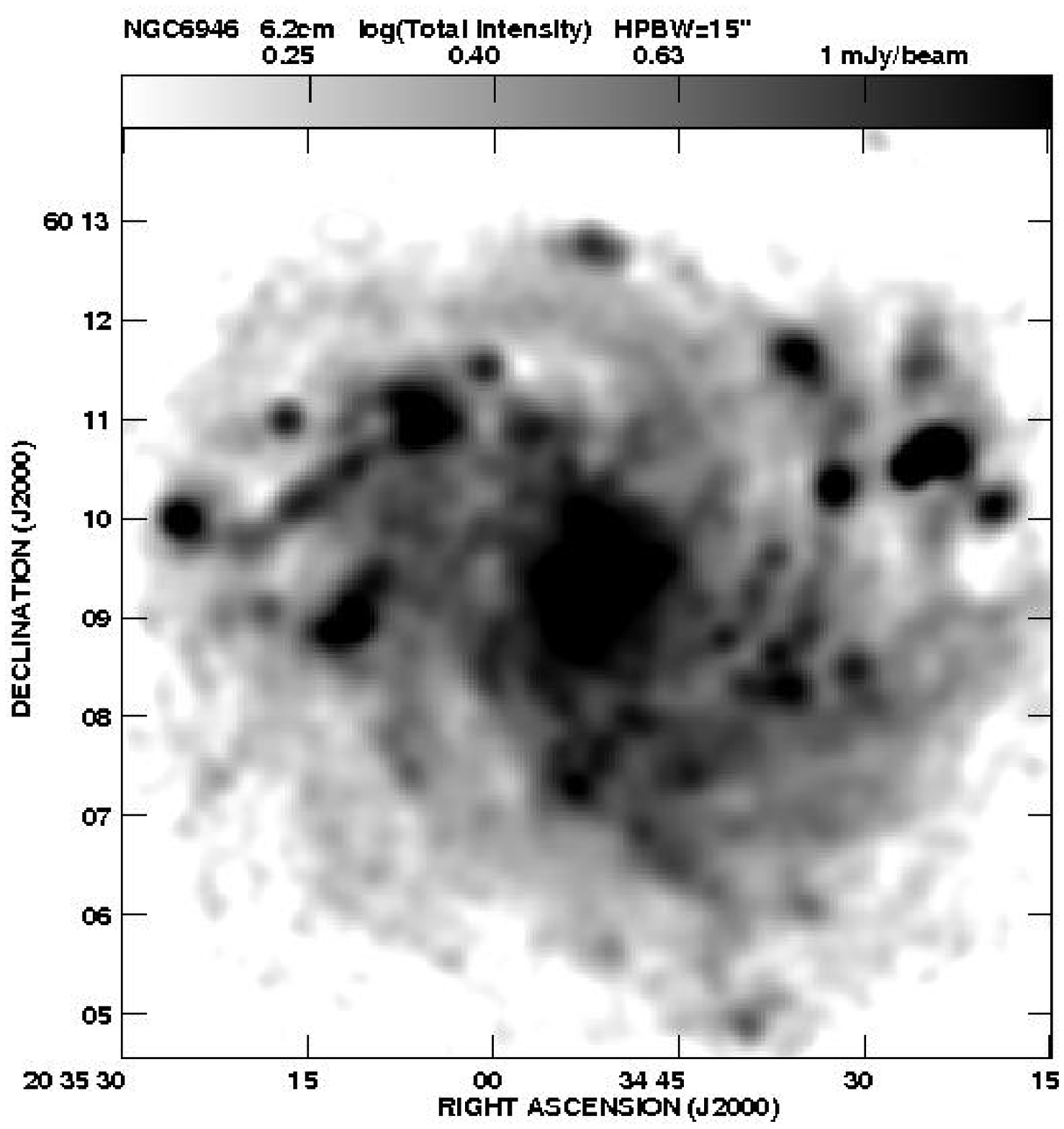}
\caption{ {\it Left:\/} Total intensity contours of NGC~6946 at
\wave{6.2} at 15\arcsec\ resolution, combined from VLA and
Effelsberg data. The contour intervals are $1, 2, 3, 4, 6, 8, 12,
16, 32, 64, 128 \times 300~\muJyb$. The background greyscale image
shows the $7\mum$ IR emission measured with the ISO satellite with
about 6\arcsec\ resolution (Dale et al.\ \cite{D00}). Here and in
the following maps, the beam size is shown in the bottom right or
left corner of each panel. {\it Right:\/} Logarithmic representation
of the total intensity at \wave{6.2} and at 15\arcsec\ resolution. }
\label{cm6i}
\end{figure*}

\begin{figure*}[htbp]
\includegraphics[bb = 62 157 528 616,width=0.475\textwidth,clip=]{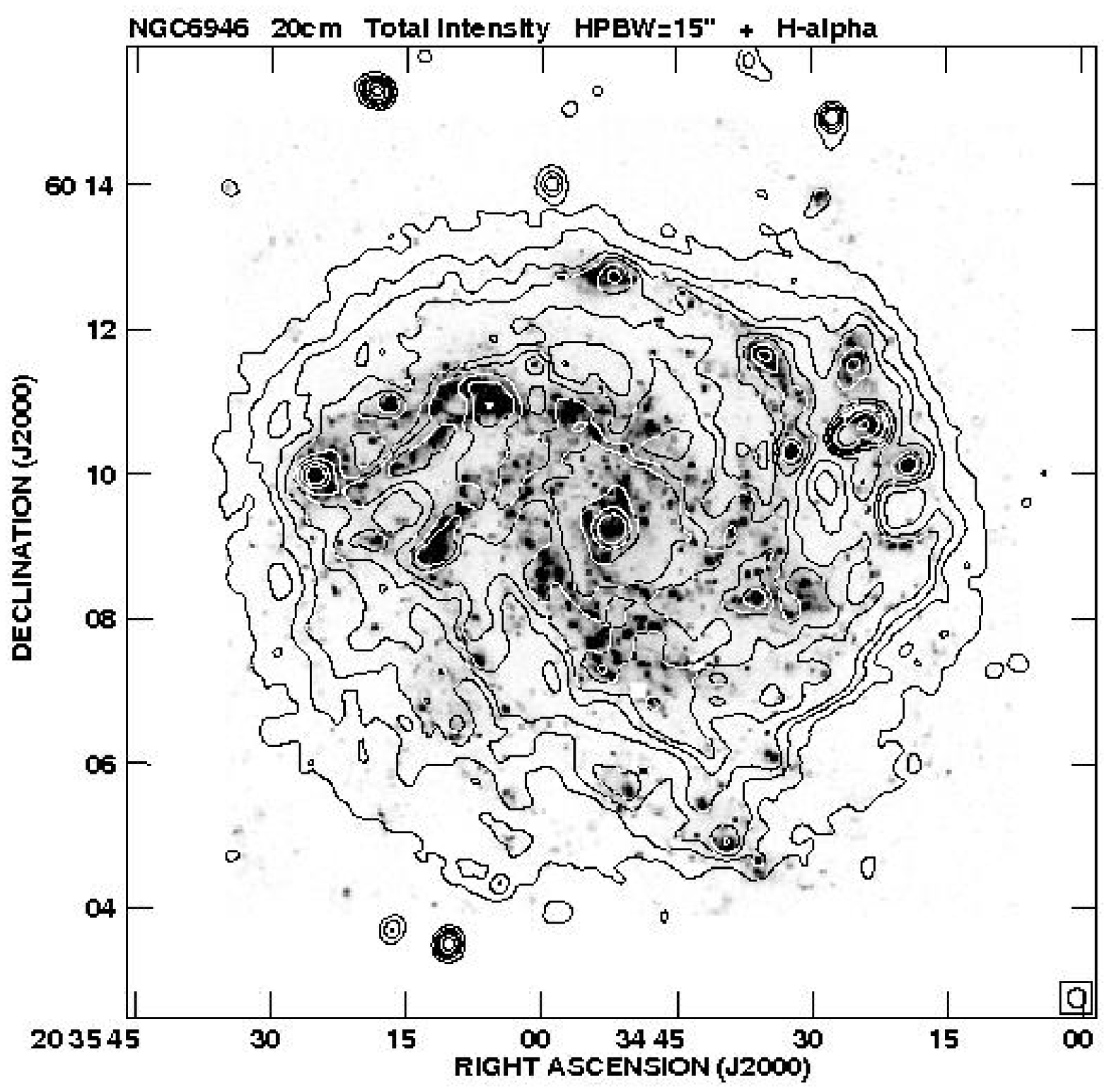}
\hfill
\includegraphics[bb = 62 142 528 601,width=0.475\textwidth,clip=]{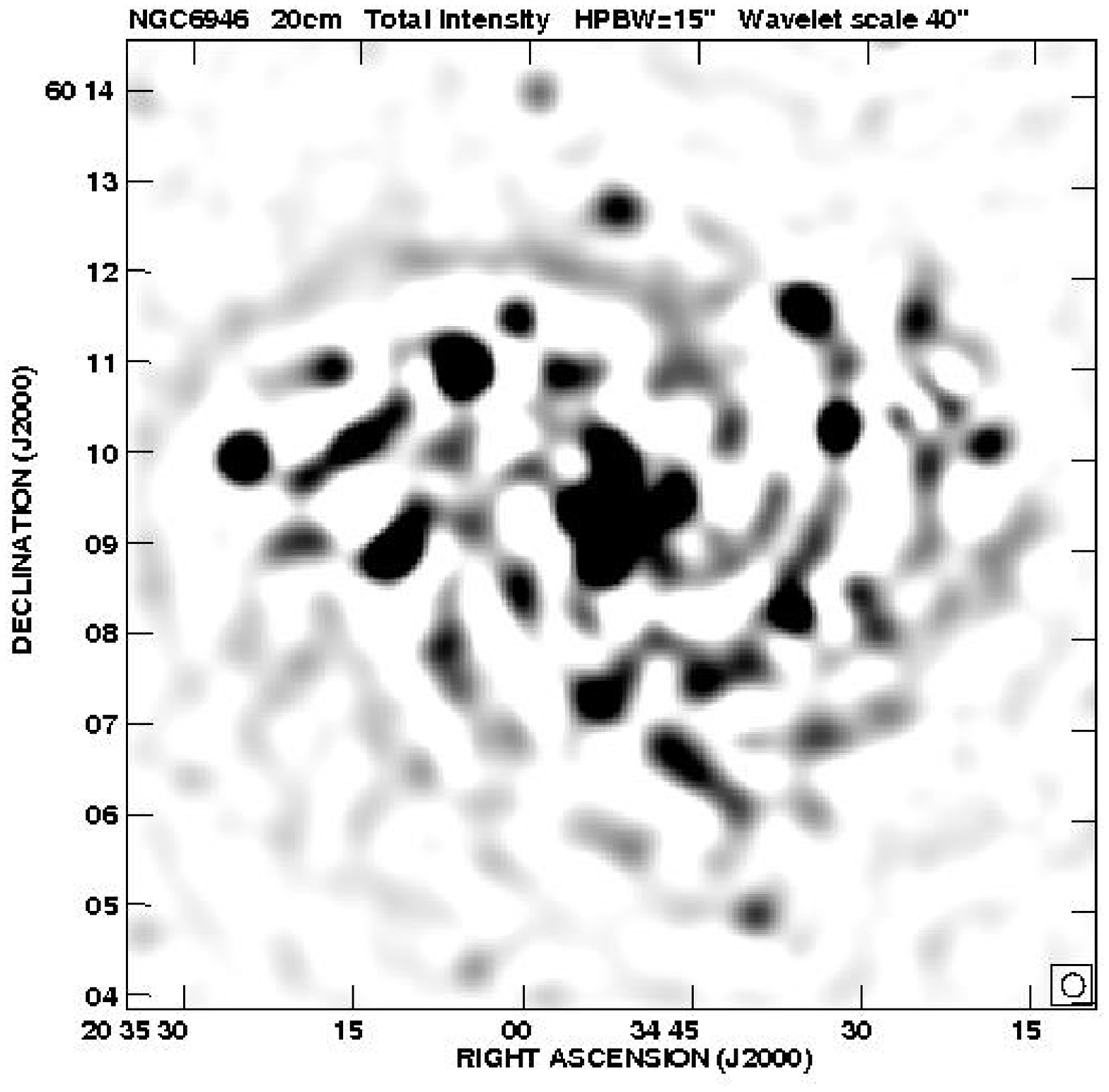}
\caption{ {\it Left:\/} Total intensity contours of NGC~6946 at
\wave{20.5} at 15\arcsec\ resolution, combined from VLA and
Effelsberg data. The contour intervals are $1, 2, 3, 4, 6, 8, 12,
16, 32, 64, 128 \times 200~\muJyb$. The background greyscale
image shows the H$\alpha$ emission (Ferguson et al.\ \cite{F98}).
{\it Right:\/} Filtered version of the \wave{20.5} image by
subtracting the three brightest sources in the fields and
transformation with an isotropic ``Mexican hat'' wavelet function
with a half-power width of 40\arcsec\ (see Frick et al.\
\cite{F01}). } \label{cm20i}
\end{figure*}

The main parameters of the radio continuum observations with the
Very Large Array (VLA) operated by the NRAO\footnote{The NRAO is a
facility of the National Science Foundation operated under
cooperative agreement by Associated Universities, Inc.} and the
Effelsberg telescope\footnote{The Effelsberg 100-m telescope is
operated by the Max-Planck-Institut f\"ur Radioastronomie in Bonn on
behalf of the Max-Planck-Gesellschaft (MPG).} are given in
Table~\ref{tab1}. Reduction of the data was performed with the
standard routines of the AIPS and NOD2 software packages,
respectively. 3C138 and 3C286 were used for calibration of flux
density and polarization angle. The secondary calibrator J2022+616
was measured during the VLA observation periods about once every 30
minutes and used for calibration of telescope gains and phases and
for correction of instrumental polarization (routine {\sc pcal}).
The accuracy of polarization angle calibration is better than
1\degr\ for both telescopes. The level of instrumental polarization
is less than 1\% at the centre of the VLA primary beam, but
increases to a few \% at the half-power radius (Condon et al.\
\cite{C98}). However, the observations presented here are obtained
from data observed over a wide range of parallactic angles so that
the instrumental polarization is smoothed out. The instrumental
polarization of the Effelsberg telescope emerges from the polarized
sidelobes with 0.3--0.5\% of the peak total intensity at the
frequencies of the observations presented in this paper.

At \wave{3.6} the VLA maps in Stokes $I$, $Q$ and $U$ were combined
(using {\sc ltess}) from three pointings centered at RA, DEC (J2000)
= $20^\mathrm{h}\ 34^\mathrm{m}\ 55\fs 9$, +60\degr\ 10\arcmin\
24\arcsec\ (north), $20^\mathrm{h}\ 34^\mathrm{m}\ 31\fs 9$, +60\degr\
09\arcmin\ 13\arcsec\ (west) and $20^\mathrm{h}\ 34^\mathrm{m}\ 52\fs
1$, +60\degr\ 06\arcmin\ 44\arcsec\ (south). Primary beam attenuation
has been corrected for. Two sets of maps were obtained by robust and
natural weighting of the $uv$ data, giving two different angular
resolutions (8\farcs 5 and 12\arcsec). The maps at 12\arcsec\
resolution were merged with the Effelsberg data at \wave{2.8} by
applying {\sc imerg} with an overlap in the baselines of the $uv$
data of 1000--1500$\lambda$. (The minimum baseline of the VLA is
about 850$\lambda$ while the maximum baseline of the Effelsberg
telescope is about 3500$\lambda$.) The small difference in
wavelengths was corrected by assuming a mean spectral index of $-0.7$,
according to Ehle \& Beck (\cite{EB93}). The errors introduced by
this assumption are negligible because the Effelsberg data are used
to fill the largest missing structures only. The combined maps were
smoothed to 15\arcsec\ ($\simeq400\pc$) resolution to increase the
signal-to-noise ratio.

At \wave{6.2} the VLA and Effelsberg maps were combined by Beck \&
Hoernes (\cite{BH96}). For this paper the original resolution of the
VLA maps of 12\farcs5\ (robust weighting) was increased to
15\arcsec\ in order to increase the signal-to-noise ratio. The
galaxy was observed again more deeply at \wave{6.2} with the
Effelsberg telescope to search for extended emission around the
galaxy.

At \wwave{18.0}{20.5} the existing D-array data by Beck (\cite{B91})
were combined with the more recent C-array data in the $uv$ plane.
Maps were generated at 15\arcsec\ resolution (robust weighting) and
corrected for primary beam attenuation. Effelsberg data were
observed at a similar wavelength. Flux integration of the VLA and
Effelsberg maps gave similar values and hence no indication for
missing large-scale structures in the VLA maps and no need for a
combination.

The maps in Stokes $Q$ and $U$ were combined to maps of linearly
polarized intensity $PI$ including the correction for positive bias
due to noise ({\sc polco} in {\sc aips} and {\sc polden} in {\sc
nod2}). Note that in the $PI$ maps the noise distribution is
non-Gaussian, and the standard deviation underestimates the
noise. Hence only rms noise values for the maps in Stokes $I$, $Q$
and $U$ are given in Table~\ref{tab1}.

\section{Results}

\subsection{Total intensity}
\label{sect:total}

The total radio intensities at \wave{6.2} and \wave{20.5} are shown
in Figs.~\ref{cm6i} and \ref{cm20i}. The emission follows closely
the distribution of IR emission, as has been discussed in previous
papers (Frick et al.\ \cite{F01}; Walsh et al.\ \cite{W02}).
According to Murphy et al. (\cite{M06}) the radio image should be a
smoothed version of the IR map due to diffusion of cosmic-ray
electrons from the star-forming regions. Figure~\ref{cm6i} does not
confirm this. The widths of several spiral arms, measured along
slices across them, are identical in the radio and IR maps to within
1\arcsec. The intensity of total synchrotron emission is
proportional to $B_\mathrm{t}^{\,1-\alpha_n}$ where $B_\mathrm{t}$ is
the component of the total magnetic field in the sky plane and
$\alpha_n$ is the synchrotron spectral index ($I_{\nu}\propto
\nu^{\,\alpha_n}$, with $\alpha_n\simeq -1$, see below). In the
equipartition case (see Sect.~\ref{sect:energies}) $I_{\nu}\propto
B_\mathrm{t}^{\,3-\alpha_n}$. As a result of this strong dependence,
the distribution of radio synchrotron emission in the disk of a spiral
galaxy like NGC~6946 is dominated by the distribution of magnetic
fields rather than by the propagation properties of cosmic rays, and
the nonthermal part of the radio-infrared correlation is due to the
coupling between magnetic fields and cold gas clouds (Niklas \& Beck\
\cite{NB97}; Hoernes et al.\ \cite{H+98}; Walsh et al.\ \cite{W02}).

NGC~6946 is surrounded by an envelope of weak radio emission which
is best visible in the most sensitive \wave{20.5} map
(Fig.~\ref{cm20i} left) and is mostly of nonthermal synchrotron
origin, as evidenced by its steep spectral index (Fig.~\ref{spec}).
Walsh et al.\ (\cite{W02}) found that the exponential scale length
of the variation of the nonthermal emission at \wave{20.5}
with galactocentric radius is $\simeq4\kpc$, about twice
that of the H$\alpha$, CO and IR emission components. In case of
energy equipartition between magnetic fields and total cosmic rays
(and constant ratio of cosmic-ray protons to electrons), the scale
lengths of the cosmic-ray electrons and of the total magnetic energy
density are $\simeq 8\kpc$, $(\frac{3-\alpha}{2}) \simeq$ 2 times longer than
that of the nonthermal emission (where $\alpha_n\simeq-1$ is the
synchrotron spectral index, see below), and the scale length of the
total magnetic field strength is $\simeq 16\kpc$, $3-\alpha_n \simeq$ 4
times longer than that of the nonthermal emission (see Fig.~\ref{en}).
If the electrons are mostly
generated in the inner disk where star formation is strongest and
lose energy when diffusing outwards, their scale length is smaller
than that obtained from the equipartition estimate while that of the
magnetic field is even larger than $16\kpc$.

In the logarithmic representation (Fig.~\ref{cm6i} right) a weak,
regular feature appears between the two northern spiral arms. The
wavelet-filtered \wave{20.5} image of scales of around 40\arcsec\
($\simeq1\kpc$) shows this feature much clearer. It is the northern
magnetic arm which is highly polarized and hence much more prominent
in polarized intensity (Sect.~\ref{sect:pol}).

An almost circular hole with very low radio emission is
observed at RA, DEC(2000) = $20^\mathrm{h}\ 34^\mathrm{m}\
20^\mathrm{s}$, 60\degr\ 09\arcmin\
40\arcsec\ (Figs.~\ref{cm6i} and \ref{cm20i}). Its diameter is about
40\arcsec\ (1\kpc). As cosmic-ray electrons can easily traverse
1\kpc\ distance, the lack of synchrotron emission must be the result
of a locally weak magnetic field. A highly regular field
perpendicular to the disk, almost along the line of sight and hence
without observable radio emission, is another explanation, though
less probable. The hole is also very prominent in $\HI$ (Braun et
al.\ \cite{B+07}) and probably represents a superbubble driven by
multiple supernova explosions which swept away the gas and magnetic
fields. $\HI$ holes are a common phenomenon in galaxies like
NGC~6946 (Kamphuis \& Sancisi\ \cite{KS93}), but the association
with a hole in radio continuum is rare.

\subsection{Thermal and nonthermal emission}
\label{sect:thermal}

\begin{figure}[htbp]
\centerline{\includegraphics[bb = 62 128 528 623,width=0.495\textwidth,clip=]{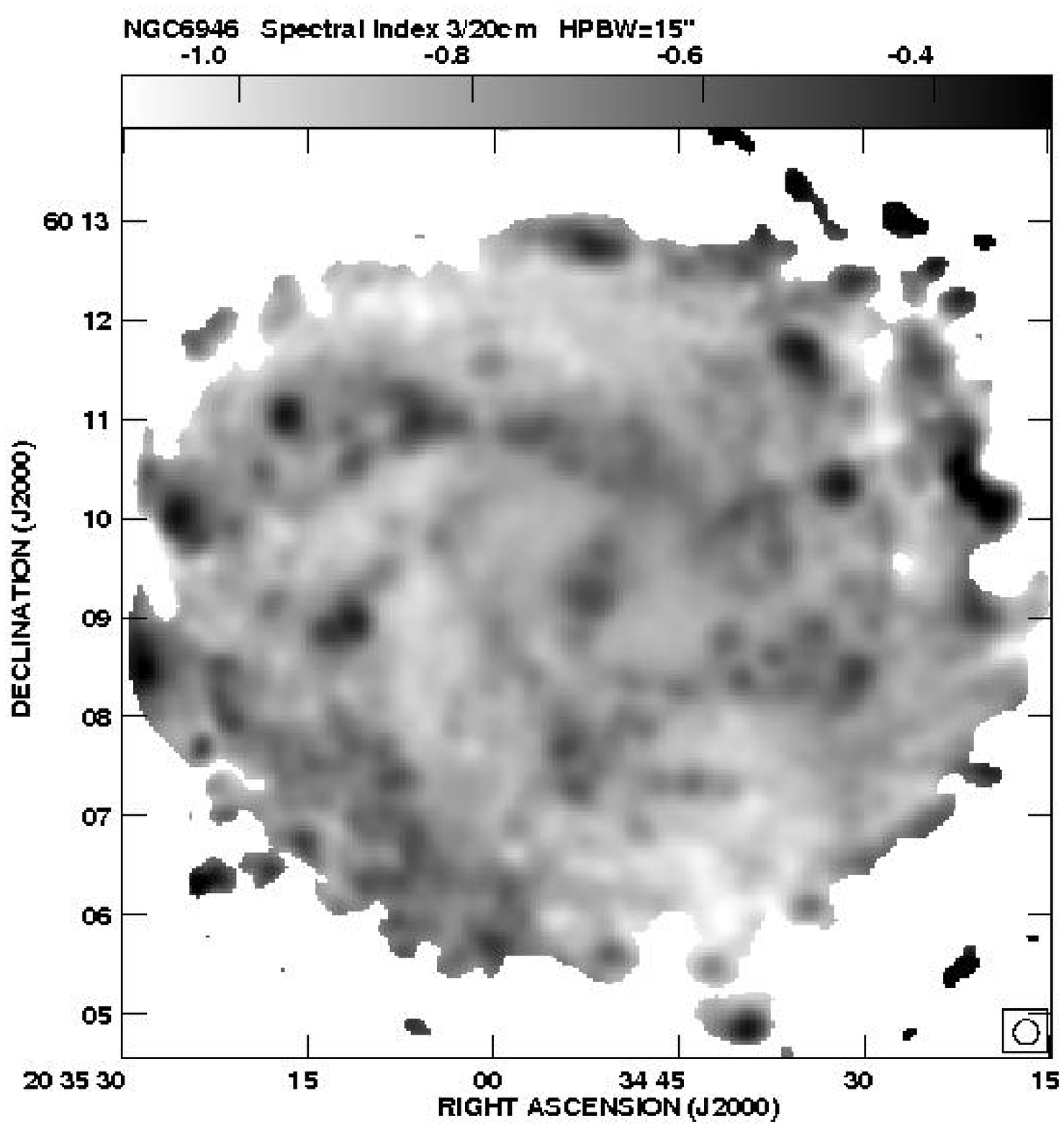}}
\caption{Spectral index distribution between \wave{20} and
\wave{3.5} at 15\arcsec\ resolution, determined for points where the
intensities at both wavelengths exceed 10$\times$ the rms noise, so
that the maximum error is $\pm0.08$.} \label{spec}
\end{figure}

The maps of intensities $I_{\nu}$ at \wave{20} and \wave{3.5} at
15\arcsec\ resolution were used to determine the radio spectral
index $\alpha$ (Fig.~\ref{spec}). The spectrum is generally flatter
in the spiral arms than outside the arms, indicating a larger
fraction of thermal emission in the arms.

\begin{figure}[htbp]
\centerline{\includegraphics[bb = 62 157 528 616,width=0.495\textwidth,clip=]{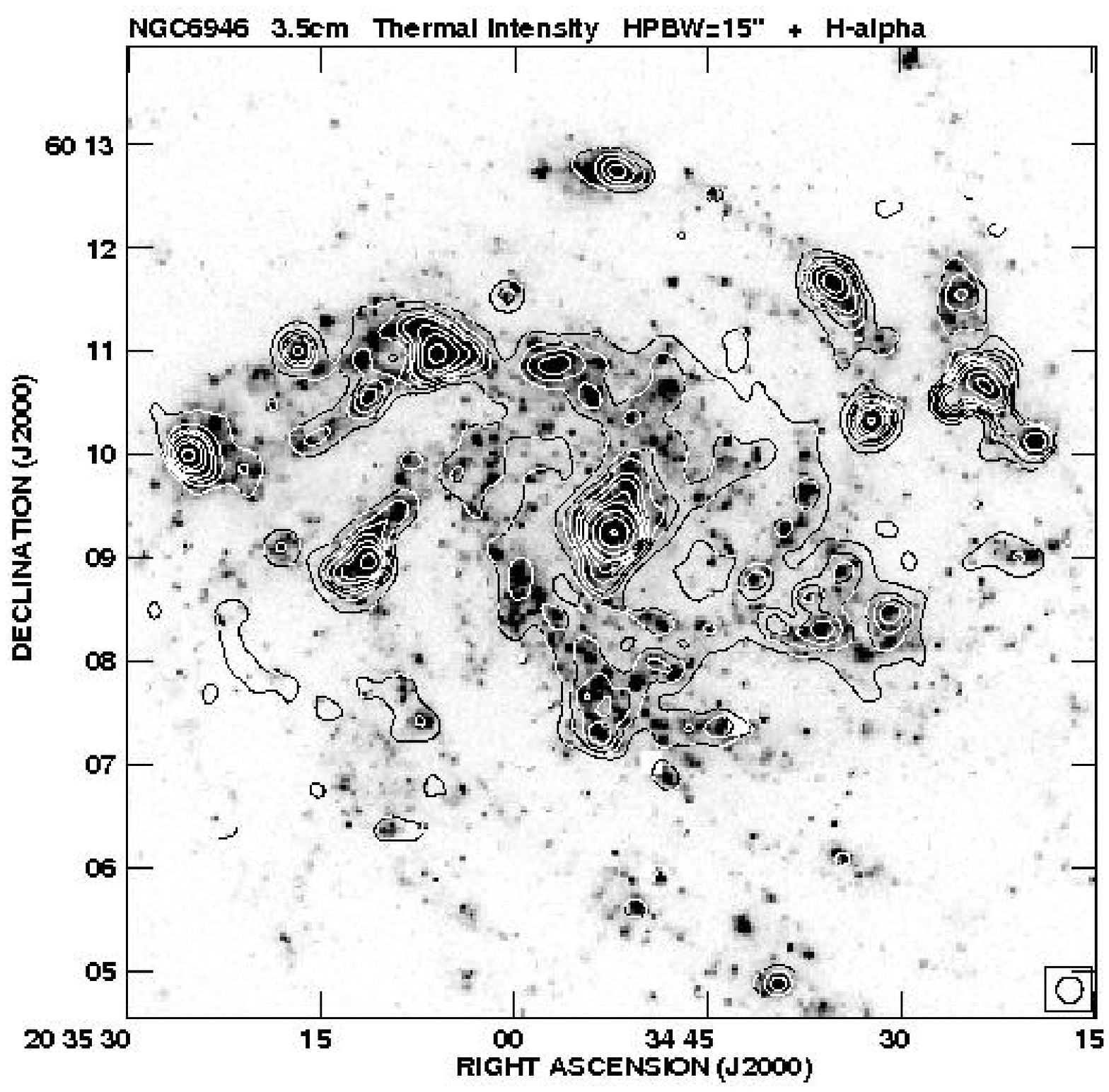}}
\caption{Contours of thermal intensity of NGC~6946 at \wave{3.5}
at 15\arcsec\ resolution.
The contour intervals are $1, 2, 3, 4, 6, 8, 12, 16, 32, 64, 128
\times 150~\muJyb$.
The background greyscale image shows the H$\alpha$ emission
(Ferguson et al.\ \cite{F98}).
}
\label{th}
\end{figure}

The thermal and nonthermal components were separated in the
classical way by assuming a constant spectral index of $-0.1$ for
the thermal and $-1.0$ for the diffuse nonthermal emission, for
which the spectral index of the outer disk was taken. The resulting
map of thermal emission agrees well with the distribution of
H$\alpha$ emission (Fig.~\ref{th}). Frick et al. (\cite{F01}) showed
that the wavelet cross-correlation between radio and optical thermal
emission is very high for scales larger than about 40\arcsec\
($\simeq1\kpc$); the weaker correlation at smaller scales is
probably due to enhanced extinction in large, dense star-forming
regions.

The assumption of a constant nonthermal spectral index
$\alpha_n=-1.0$ is clearly too simplistic. Firstly, the regions of
high star-formation rate in the spiral arms contain a large fraction
of young cosmic-ray electrons released from supernova remnants which
have a flat spectrum of $\alpha_n\simeq-0.6$ (Beck \& Krause\
\cite{BK05}). Away from the spiral arms electrons lose energy by
synchrotron, inverse Compton and escape losses, all of which are
energy dependent and hence steepen the spectrum. Secondly, the
spiral arms contain a contribution of direct synchrotron emission
from supernova remnants which have spectral indices $\alpha_n$
between $0$ and $-0.9$ (Green\ \cite{G01}). Both effects lead to an
\emph{overestimate} of the thermal emission in the spiral arms and
hence an \emph{underestimate} of the nonthermal emission. More
reliable estimates of the thermal radio emission require
extinction-corrected H$\alpha$ maps (Tabatabaei et al.\
\cite{T+07}).

\subsection{Magnetic field strengths and energy densities}
\label{sect:energies}

The strengths of the total magnetic field $B_t$ and its
regular\footnote{Note that $B$-vectors of linearly polarized
emission can trace either {\em regular}\ magnetic
fields (i.e. preserving their direction within the telescope beam,
also called coherent fields) or {\em anisotropic}\
fields (i.e. with multiple field reversals within the beam,
also called incoherent fields). To distinguish between these
two components, additional Faraday rotation data is needed.
The fields observed in polarization are called ``ordered''
throughout this paper. If Faraday rotation data are available,
``regular'' or ``anisotropic'' are used. -- Also note that
the equipartition strengths of the ordered field are always lower
limits due to the limited resolution (beam depolarization) and
Faraday depolarization effects.} component $B_\mathrm{reg}$ in the sky
plane can be derived from the total synchrotron intensity and its
degree of linear polarization (see Sect.~\ref{sect:pol}), assuming
equipartition between the energy densities of the magnetic field and
the total cosmic rays, a value for the ratio $R$ between the number
densities of cosmic-ray protons and electrons, the pathlength $L$
through the synchrotron-emitting medium and the synchrotron spectral
index $\alpha_n$ (Beck \& Krause\ \cite{BK05}). With $R=100$,
$L=1\kpc$ and $\alpha_n=-1$, the average total field strength varies
from $25\muG$ in the inner bar to $15\muG$ in a ring at 5\kpc\
($\simeq3\arcmin$) radius of 0.5\kpc\ width, the average
ordered field from $8\muG$ near the centre to $6\muG$ in the
same ring at 5\kpc\ radius. Typical values of the total field
strength in the main optical spiral arms are 18--20$\muG$, those of
the ordered field $\le5\muG$. The strengths of the ordered,
mostly regular field in the two main magnetic arms (which
are strongest at around 5\kpc\ radius) is typically 8--10$\muG$.

The faintest total intensity of $\simeq1\mJy$ per 147\arcsec\ beam
detected in the outer galaxy with the Effelsberg telescope at
\wave{6.2} is near the confusion limit of unresolved background
sources of $0.5\mJyb$ with this telescope (Reich, priv. comm.) and
gives a detection limit for total (equipartition) fields of about
$6\muG$. The Effelsberg total intensity limit corresponds to
$\simeq10\muJy$ per 15\arcsec\ beam, which is below the rms noise of
the VLA maps in this paper. The faintest polarized intensity
(Fig.~\ref{pol6}) of $\simeq200\muJy$ per 147\arcsec\ beam
(corresponding to $\simeq2\muJy$ per 15\arcsec\ beam) gives a
detection limit for ordered fields of about $3\muG$.
Detection of weaker total or ordered fields in galaxies
would need unrealistically deep integrations. Detection of weaker
regular fields by their intrinsic Faraday rotation of
polarized emission from background sources is possible but
also requires deep integrations to obtain a sufficiently large
number of background sources. Hence, the data of NGC~6946 presented
in this paper demonstrate the limits of present-day telescopes.

\begin{figure}[htbp]
\centerline{\includegraphics[bb = 69 349 507 645,width=0.495\textwidth,clip=]{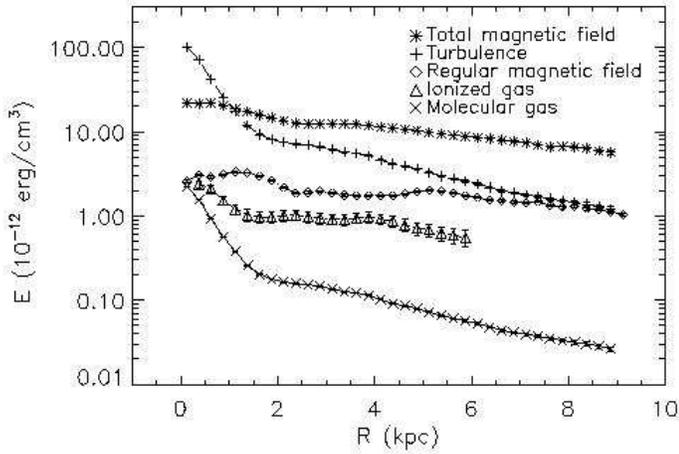}}
\caption{Energy densities and their radial variations in the spiral
galaxy NGC~6946.} \label{en}
\end{figure}

The relative importance of various competing forces in the
interstellar medium can be estimated by comparing the corresponding
energy densities or pressures. In the local Milky Way, the
average pressures of turbulent gas motions, cosmic rays and
magnetic fields are similar ($\simeq 1 \cdot 10^{-12}$~erg~cm$^{-3}$),
while the thermal pressure of the warm gas is smaller by a
factor of 3--5 (Boulares \& Cox\ \cite{BC90}; Cox\ \cite{C05}).
In numerical MHD simulations, much of the gas
falls off the equipartition lines between local thermal and magnetic
pressures as well as between local thermal and kinetic pressures,
and even the average values do not agree (Avillez \& Breitschwerdt\
\cite{AB05}; Mac Low et al.\ \cite{MB05}). On the other hand, these
simulations do not include dynamo action which may help to balance
magnetic and kinetic energies. The new data available for NGC~6946
allow a study of average pressures as a function of galacto-centric
radius (Fig.~\ref{en}), results which are hard to achieve for the
Milky Way.

The energy density of the total equipartition magnetic field
($B^2_\mathrm{t}/8\pi$) in the disk, identical to that of the total
cosmic rays, was derived from the map of nonthermal synchrotron
intensity, assuming a constant pathlength through the emitting
medium of 1\kpc. The energy density of the ordered magnetic
field was derived from the map of polarized intensity. The thermal
energy density (${3\over2} \langle n_\mathrm{e} \rangle k T$) of the
warm ionized gas ($T\simeq 10^4$~K) was calculated
from the map of thermal radio emission
(giving the emission measure $EM = L \langle n_\mathrm{e}^2
\rangle$ which is dominated by $\HII$ regions, see Fig.~\ref{th}),
using a pathlength $L$ of 1\kpc, a constant volume filling factor
$f$ of 5\% (Ehle \& Beck\ \cite{EB93}). The average number density
$\langle n_\mathrm{e} \rangle = \sqrt{f \langle
n_\mathrm{e}^2 \rangle}$ of the $\HII$ regions decreases from about
$0.5\ccm$ at 1\kpc\ radius to about $0.3\ccm$ at 5\kpc\ radius.
The average number density $\langle n \rangle$ of the total
neutral gas (molecular + atomic) was determined from the CO map of
Walsh et al.\ (\cite{W02}) and the $\HI$ map of Kamphuis \& Sancisi
(\cite{KS93}), assuming for simplicity a constant scale height of
the disk of 100\pc. $\langle n \rangle$ decreases from about
$50\ccm$ at 1\kpc\ radius to about $7\ccm$ at 5\kpc\ radius.
The molecular gas traced by the CO emission dominates until
5\kpc\ radius. Assuming a constant temperature of $T=50$~K, the
energy density of the molecular gas was computed and is shown in
Fig.~\ref{en}. The contributions of the cold and warm atomic gas
cannot be determined with the existing low-resolution data, so that
the thermal energy density of the total neutral gas is unknown in
NGC~6946. To compute the kinetic energy density (${1\over2} \rho
v^2$) of the turbulent motion of the total neutral gas, its
turbulent velocity was assumed to be $v_\mathrm{turb}=7\kms$, the
typical value derived for the neutral gas in our Galaxy
(Boulares \& Cox\ \cite{BC90}; Kalberla \& Kerp\ \cite{KK98})
and from ISM models (Fletcher \& Shukurov\ \cite{FS01}).

The energy densities of all components in NGC~6946 are about one order
of magnitude larger than in the Milky Way (Cox\ \cite{C05}).
The energy density of the ionized gas $E_\mathrm{th}$ in NGC~6946
(Fig.~\ref{en}) is about one order of magnitude smaller than that of
the magnetic field $E_\mathrm{magn}$. This result is similar to that
derived for the Milky Way (Cox\ \cite{C05}). The ISM seems to be a
low-$\beta$ plasma ($\beta=E_\mathrm{th}/E_\mathrm{magn}$). 3-D
magnetohydrodynamical models for the ISM predicted a large variation
of local $\beta$ values, with the average value steeply increasing
with temperature (Mac Low et al.\ \cite{MB05}; Avillez \&
Breitschwerdt\ \cite{AB05}).

The thermal energy density of the ionized gas $E_\mathrm{th}$ in
Fig.~\ref{en} and hence $\beta$ may be underestimated if there is a
significant contribution from hot gas, or if the
diffuse warm ionized gas dominates over the $\HII$ regions, so
that the effective filling factor $f$ of the warm ionized gas would
be larger than 5\% ($\langle n_\mathrm{e} \rangle \propto
\sqrt{I_\mathrm{th} \, f}$). However, the values of $\langle
n_\mathrm{e} \rangle$ for the $\HII$ regions in NGC~6946 (see above)
are more than one order of magnitude larger than for the warm
diffuse ionized gas (Sect.~\ref{sect:rm}) and about two orders of
magnitude larger than typical values for the hot gas (Fletcher \&
Shukurov\ \cite{FS01}). Finally, $E_\mathrm{th}$ could be
twice larger if the contribution of protons is also
considered. In summary, no major uncertainties in $E_\mathrm{th}$
are expected.

The large energy density of the turbulent motions compared to
the thermal energy density in Fig.~\ref{en} means that turbulence in
the diffuse ISM is supersonic, which is consistent with 3-D MHD
models for the ISM (Avillez \& Breitschwerdt\ \cite{AB05}). While
supersonic turbulence in molecular gas clouds of the Milky Way is
observed and helps to support the clouds (Mac Low \& Klessen\
\cite{M04}), supersonic turbulence in the diffuse ISM would lead to
widespread shocks and hence is hard to maintain. The results
presented here are puzzling and need further investigations in other
galaxies.

In the inner parts of NGC~6946 the energy densities of the total
magnetic field and turbulent gas motions are similar, while the
field dominates in the outer parts. The reason is the large radial
scale length of the total magnetic energy ($7.0\pm0.1\kpc$ for radii
larger than 3\kpc) compared to the scale length of about 3\kpc\ of
the neutral gas density (Walsh et al.\ \cite{W02}). This is in
apparent conflict with the generation of interstellar
magnetic fields from turbulent gas motions. Before
considering physical explanations, like radial diffusion of the
magnetic field (Priklonsky et al.\ \cite{P00}), field configurations
in a wind-driven halo (Breitschwerdt et al.\ \cite{BD02}), or a
supra-equipartition turbulent dynamo (Belyanin et al.\ \cite{BS93}),
possible bias effects of the values shown in Fig.~\ref{en}
have to be discussed.

The discrepancy between total magnetic and turbulent energy
densities may indicate that the turbulent velocity of the neutral
gas in NGC~6946 is underestimated. $6\kms$ and $9\kms$ of the cold
and warm neutral gas components, respectively, are needed for the
vertical hydrostatic equilibrium, while the total turbulent velocity
is $\sqrt{3}$ times higher (Fletcher \& Shukurov\ \cite{FS01}).
Another increase could be caused by the high star-formation rate in
NGC~6946. As a result, the magnetic and turbulent energy densities
could become similar until several kpc radius. In the outer parts of
the galaxy, turbulence generated by the magneto-rotational
instability (MRI) (Sellwood \& Balbus\ \cite{S99}; Dziourkevitch et
al.\ \cite{D04}) could be stronger than turbulence excited by star
formation and increase the turbulent energy density and scale
length.

The discrepancy between total magnetic and turbulence energy
densities for radii larger than about 2\kpc\ may also indicate that
the magnetic field strength or the scale length of magnetic energy
density are overestimated. If energy equipartition between magnetic
fields and cosmic rays does not hold, the estimates of field
strength based on this assumption would be invalid. If, e.g.,
the density of cosmic-ray electrons is constant in the galaxy, the scale
length of the magnetic energy density would decrease to 3.5\kpc,
similar to the scale length of the energy density of turbulence.
However, the sources of cosmic rays are believed to be related to
star-forming regions, which are concentrated in the inner galaxy,
and the propagation of cosmic-ray electrons is limited by
synchrotron losses, so that a constant density of cosmic-ray
electrons can hardly be maintained. $\gamma$-ray data from the Milky
Way indeed indicate a radial decrease of the cosmic ray distribution
(Strong \& Mattox\ \cite{SM96}). The aging of cosmic-ray electrons
towards the outer galaxy leads to a steeper decrease and hence to an
increasing proton-to-electron ratio $R$ and to an {\em
underestimate} of the magnetic energy as shown in Fig.~\ref{en}.
This would enhance the discrepancy with the turbulent energy.

Fluctuations of the total field strength across the telescope
beam or along the line of sight are a possible source of uncertainty
of the equipartition estimate because the observed synchrotron
emission traces $(\langle B_\mathrm{t}^4 \rangle)^{1/4}$ (Beck\ \&
Krause\ \cite{BK05}, for a synchrotron spectral index $\alpha_n=-1$)
and hence is biased towards regions with strong fields. With
fluctuations in strength of amplitude $a$ (where $a=\delta
B_\mathrm{t}/\langle B_\mathrm{t} \rangle$) the total equipartition
strength is too large by a factor of $(1+6a^2+a^4)^{1/4}$. Note that
$\delta B_\mathrm{t}$ is smaller than the random field $B_\mathrm{r}$ as
derived from Faraday depolarization (Sect.~\ref{sect:dp}) because
fluctuations of the field direction do not contribute to $\delta
B_\mathrm{t}$. For the same reason, the degree of field regularity
(Fig.~\ref{reg}) also cannot be used to estimate $a$. Regarding
$a\simeq 1$ as the upper limit, the maximum overestimate is 1.7 times
for the total field strength and 2.8 for the total magnetic energy. In
summary, there may be room to argue for an overestimate of the total
magnetic energy density in Fig.~\ref{en} by a factor of about 2--3.

The energy density of the ordered field in NGC~6946 is 5--8
times smaller than that of the total field, while the scale length
is similar, $8.2\pm0.8\kpc$ for radii larger than 6\kpc.
Ordered fields are detected in NGC~6946 up to about 15\kpc\
radius (Fig.~\ref{pol6}) and regular fields up to 12\kpc\
radius (Sect.~\ref{sect:rm}), but these may extend much
further, to at least 20\kpc\ (12\farcm5) radius until which $\HI$
emission is detected (Boomsma et al.\ \cite{BO05};
Braun et al.\ \cite{B+07}, Fig.~\ref{HI}). A
deep survey of Faraday rotation of polarized background sources is
needed to search for magnetic fields at large radii.

If the slow decrease of the magnetic field energy density in
Fig.~\ref{en} continues to even larger radii, it may
affect the rotation
curve in the outer galaxy, as proposed by Battaner \& Florido
(\cite{BF00}). According to Battaner \& Florido (\cite{BF07})
the flat radial profile of the magnetic energy in NGC~6946 causes a
net inward force.

\subsection{Polarized intensity}
\label{sect:pol}

\begin{figure*}[htbp]
\includegraphics[bb = 62 157 528 616,width=0.475\textwidth,clip=]{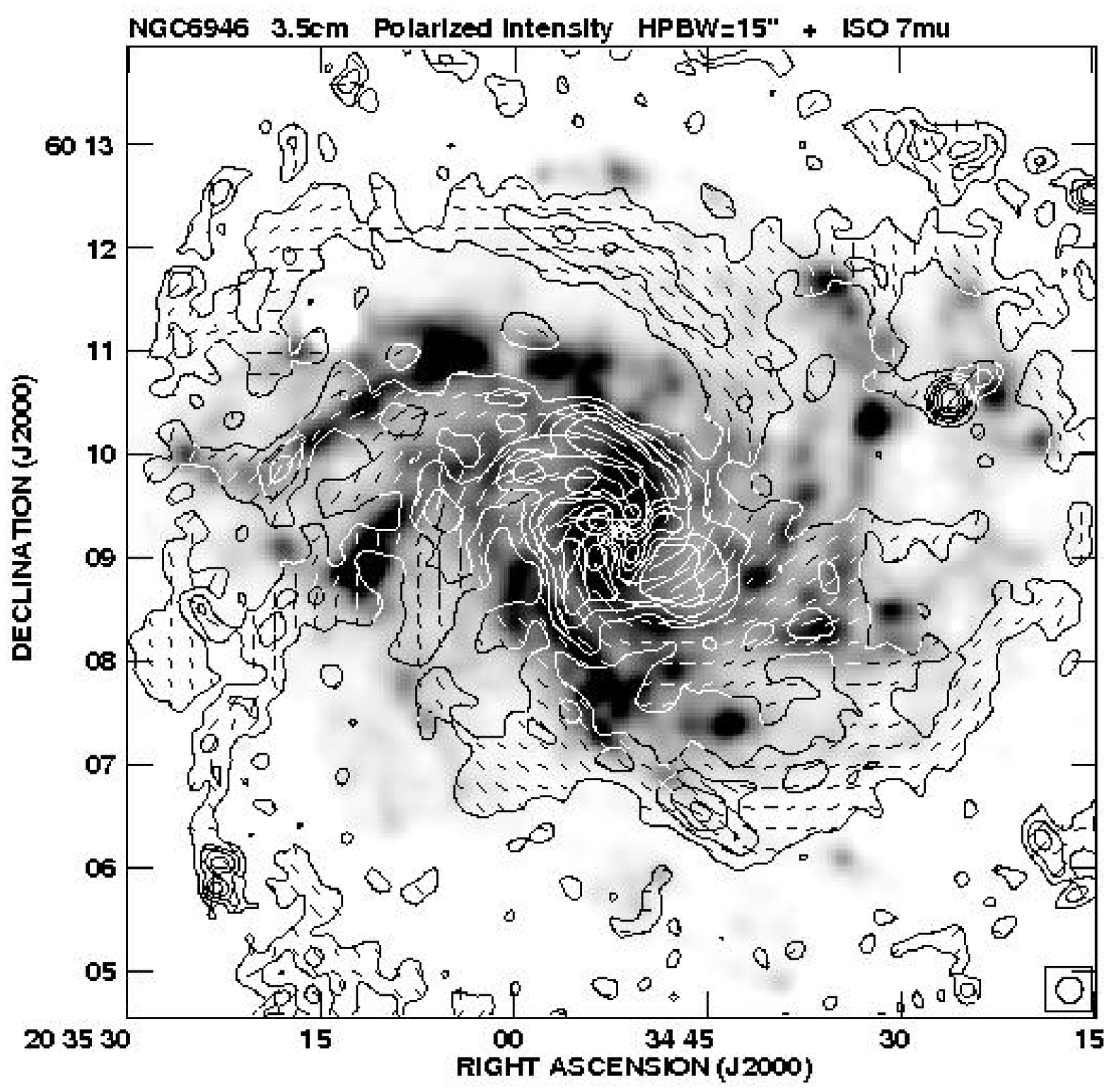}
\hfill
\includegraphics[bb = 62 157 528 616,width=0.475\textwidth,clip=]{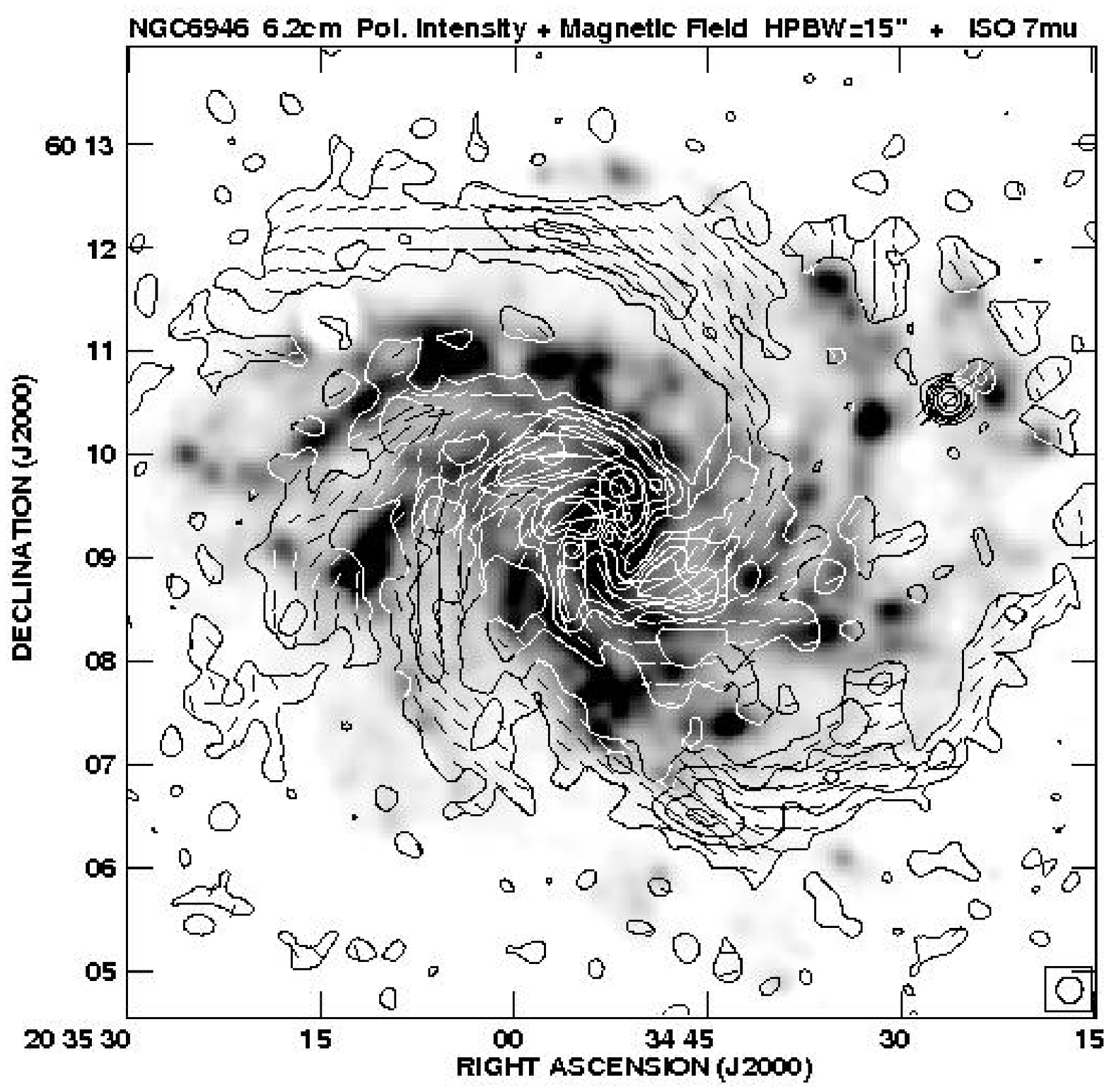}
\vspace*{0.2cm}
\includegraphics[bb = 62 157 528 616,width=0.475\textwidth,clip=]{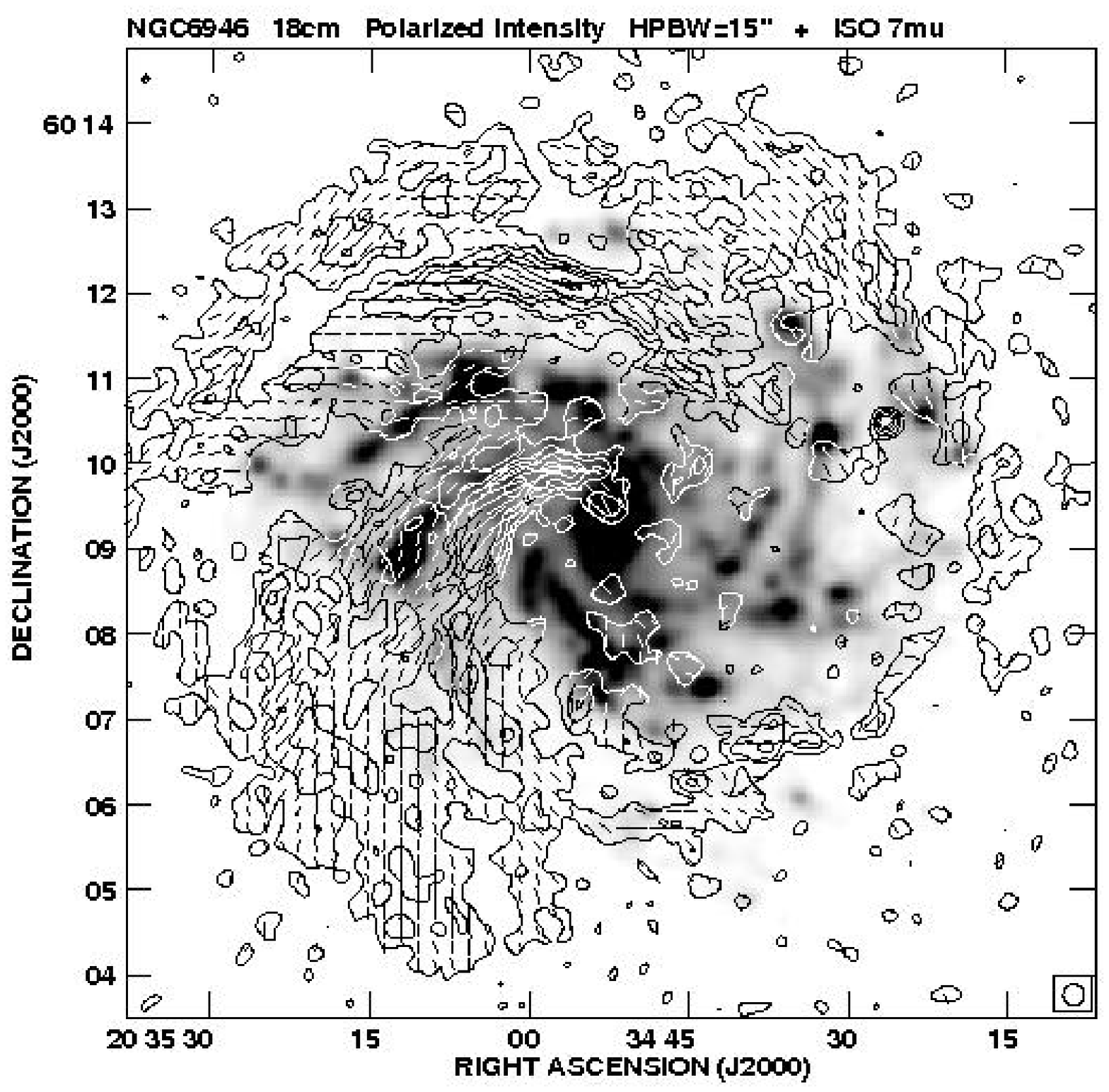}
\hfill
\includegraphics[bb = 62 157 528 616,width=0.475\textwidth,clip=]{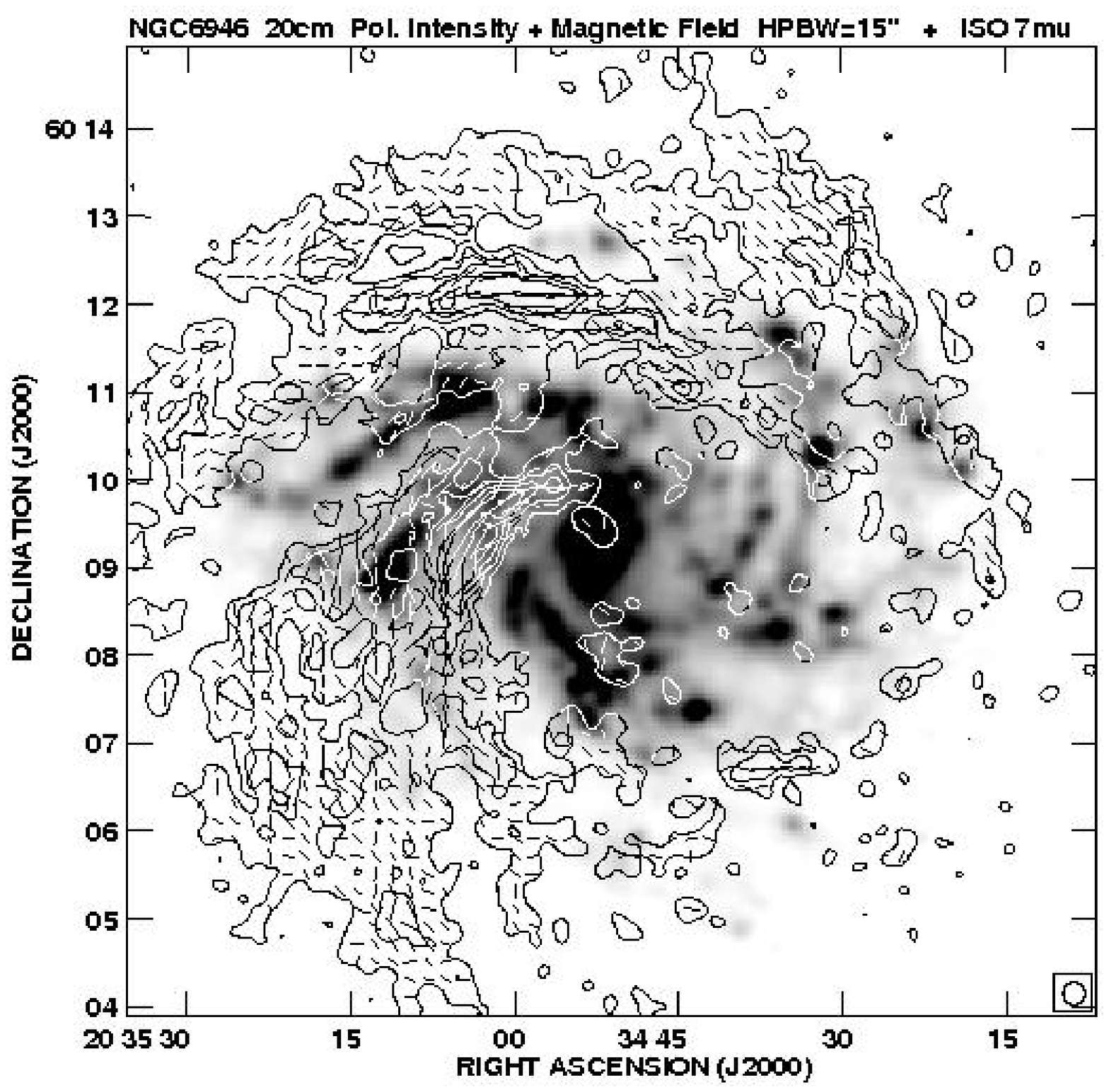}
\caption{
{\it Top left:\/}
Polarized intensity contours and observed
$B$-vectors ($E$+90\degr) of NGC~6946 at \wave{3.5}
at 15\arcsec\ resolution, combined from VLA and Effelsberg data.
The contour intervals are $1, 2, 3, 4, 6, 8 \times 40\muJyb$.
The vector length is proportional to polarized intensity,
12\arcsec\ length corresponds to 100~$\muJyb$.
Faraday rotation is low ($\le$10\degr) at this wavelength.
The background greyscale image shows the $7\mum$ IR emission measured
with the ISO satellite with about 6\arcsec\ resolution (Dale et al.\
\cite{D00}).
{\it Top right:\/}
Polarized intensity contours and $B$-vectors at
\wave{6.2}, corrected for Faraday rotation (see Fig.~\ref{rm} left)
at 15\arcsec\ resolution, combined from VLA and Effelsberg data.
The contour intervals are $1, 2, 3, 4, 6, 8 \times 60\muJyb$.
The vector length is proportional to polarized intensity,
12\arcsec\ length corresponds to 100~$\muJyb$.
{\it Bottom left:\/}
Polarized intensity contours and observed $E$-vectors at \wave{18.0},
not corrected for Faraday rotation, at 15\arcsec\ resolution, combined
from VLA C- and D-array data.
The contour intervals are $1, 2, 3, 4, 6 \times 50\muJyb$.
The vector length is proportional to
polarized intensity, 12\arcsec\ length corresponds to 100~$\muJyb$.
{\it Bottom right:\/}
Polarized intensity contours and $B$-vectors at \wave{20.5},
corrected for Faraday rotation (see Fig.~\ref{rm} right),
at 15\arcsec\ resolution, combined from VLA C- and D-array data.
The contour intervals are $1, 2, 3, 4, 6 \times 50\muJyb$.
The vector length is proportional to
polarized intensity, 12\arcsec\ length corresponds to 100~$\muJyb$.
}
\label{pol}
\end{figure*}

\begin{figure}[htbp]
\centerline{\includegraphics[bb = 62 149 528 616,width=0.495\textwidth,clip=]{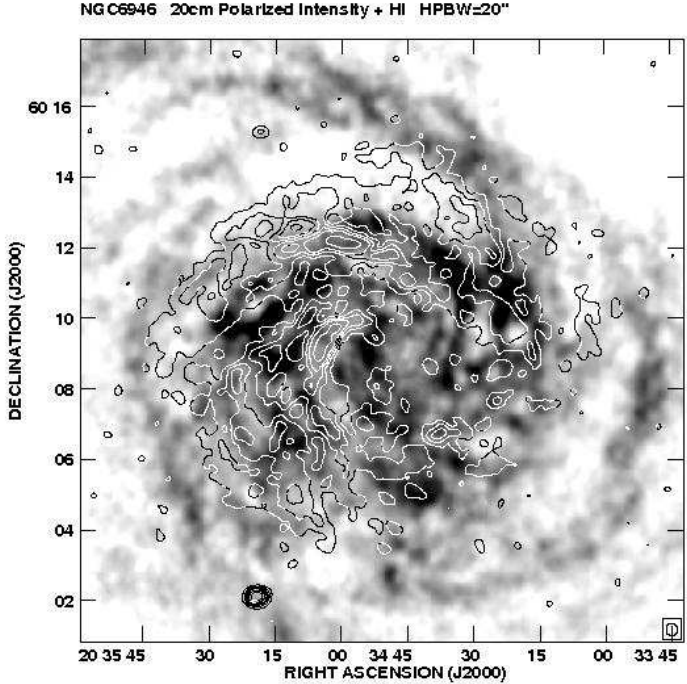}}
\caption{Polarized intensity contours of NGC~6946 at \wave{20.5} at
20\arcsec\ resolution, observed with VLA. The contour intervals are
$1, 2, 3, 4, 6, 8~\times 60~\muJyb$. The background greyscale image
shows the $\HI$ emission measured with the Westerbork telescope at
about the same resolution (Boomsma et al.\ \cite{BO05};
Braun et al.\ \cite{B+07}).} \label{HI}
\end{figure}

\begin{figure}[htbp]
\centerline{\includegraphics[bb = 62 178 528 609,width=0.495\textwidth,clip=]{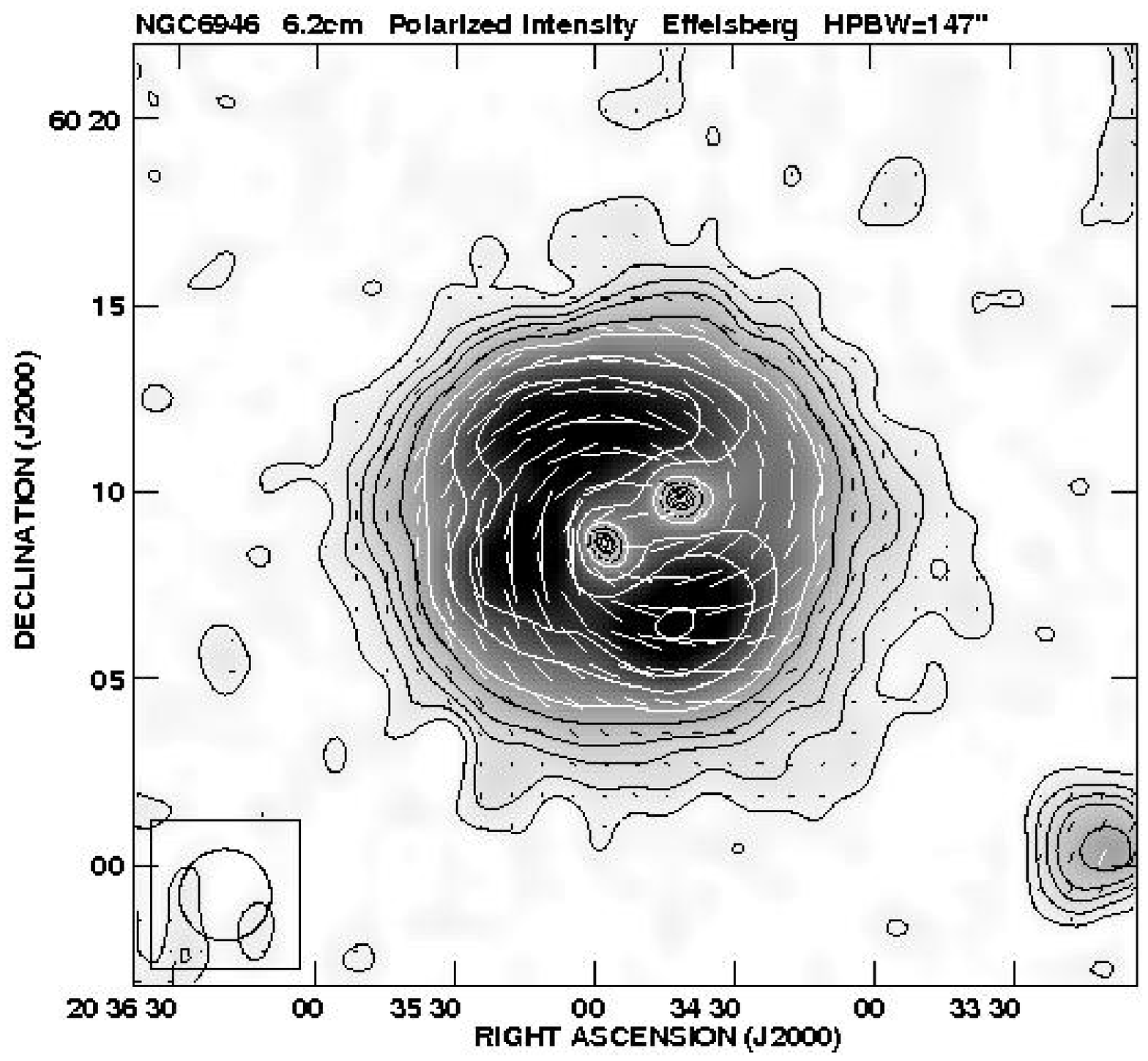}}
\caption{Polarized intensity contours and $B$-vectors ($E$+90\degr)
of NGC~6946 at \wave{6.2} at 2\farcm45\ resolution, observed with
the Effelsberg telescope. The contour intervals are $1, 2, 3, 4, 6,
8, 12, 16, 24~\times 200~\muJyb$. The vector length is proportional
to polarized intensity, 20\arcsec\ length corresponds to
1000~$\muJyb$. The vector orientations are not corrected for Faraday
rotation. } \label{pol6}
\end{figure}

\begin{figure}[htbp]
\centerline{\includegraphics[bb = 62 178 536 609,width=0.495\textwidth,clip=]{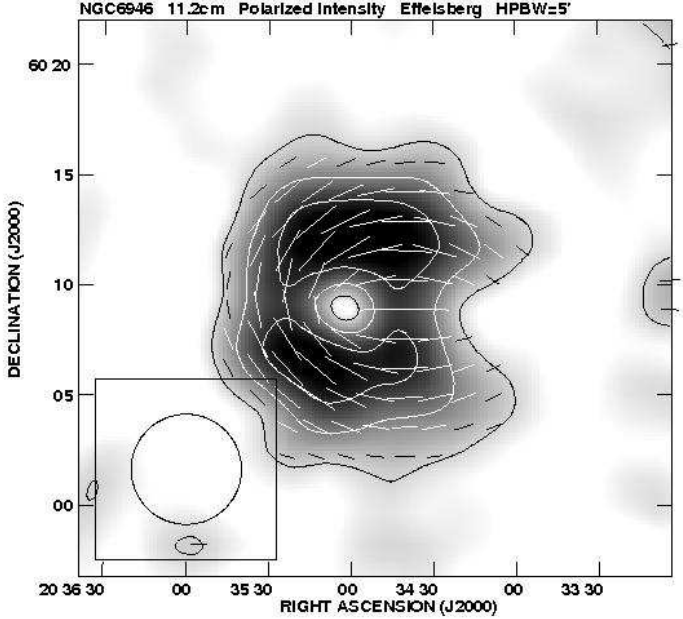}}
\caption{Polarized intensity contours and $B$-vectors ($E$+90\degr)
of NGC~6946 at \wave{11.2} at 5\arcmin\ resolution, observed with
the Effelsberg telescope. The contour intervals are $1, 2, 3 \times
2000~\muJyb$. The vector length is proportional to polarized
intensity, 1\arcmin\ length corresponds to 3000~$\muJyb$. The vector
orientations are not corrected for Faraday rotation. } \label{pol11}
\end{figure}

Figs.~\ref{pol}--\ref{pol3} show the distributions of the
polarized emission at the five observation frequencies. There is
very little similarity to the maps of total emission. The polarized
emission in the optical spiral arms is low; it is anticorrelated
with the H$\alpha$ emission (Frick et al.\ \cite{F01}). The
fractional polarization in the inner optical arms of 1--5\% at
\wwave{3}{6} wavelengths is low due to wavelength-independent
depolarization by unresolved field structures, either fields tangled
by star-formation activity or by turbulent fields. In the outer arms
the fractional polarization increases to 5--10\%. At \wwave{18}{20}
the fractional polarization in the arms is smaller than 2\%.

Note that the orientations of the observed $E$-vectors at
\wave{18.0} (Fig.~\ref{pol} bottom left) and \wave{20.5} are by
chance similar to those of the magnetic field because Faraday
rotation at these wavelengths originates mainly in the Galactic
foreground which has $RM_\mathrm{fg}\simeq40\radm$, giving 74\degr\ at
\wave{18.0} and 96\degr\ at \wave{20.5} (see Sect.~\ref{sect:rm}).

The Faraday-corrected magnetic field vectors (Fig.~\ref{pol} right)
appear to be less ordered than the observed $B$-vectors
(Fig.~\ref{pol} left). This is due to the relatively large errors in
Faraday rotation which had to be determined from data over a small
frequency range, leading to errors in the field orientation of up to
15\degr\ between \wave{3} and \wave{6} and up to 45\degr\ between
\wave{18} and \wave{20} at the cutoff limit of 4 times the rms noise
(Sect.~\ref{sect:rm}).

The most striking phenomena in Fig.~\ref{pol} are the two bright
\emph{magnetic arms}\ between the optical arms, known already from
the previous \wave{6.2} observations (Beck \& Hoernes\ \cite{BH96}).
The magnetic arms are visible at all wavelengths, except in the
single-dish Effelsberg maps (Figs.~\ref{pol6} and \ref{pol11}) where
the beam is too large to resolve them.

The fractional polarization $p$ observed at \wave{3} and \wave{6} is
30--45\% in the northern magnetic arm and 20--30\% in the southern
magnetic arm. These are among the highest degrees of polarization
observed in any spiral galaxy so far, similarly high as in the
emission ring of M~31 (Berkhuijsen et al.\ \cite{BB03}). The
fractional polarization of the nonthermal emission $p_n$ is obtained
by subtraction of the unpolarized thermal contribution. As the
thermal emission is small in the interarm regions (Fig.~\ref{th})
$p_n$ is only marginally higher than $p$ in the magnetic arms.
Subtraction of the smooth, mostly unpolarized background from the
total intensity increases $p$ in the northern arm to about 75\% at
\wave{3} and \wave{6}, which is the maximum possible degree of
linear polarization $p_0$ of $p_0=(3-3\alpha_n)/(5-3\alpha_n)$
(where $\alpha_n$ is the nonthermal spectral index). \emph{The
magnetic field in the northern magnetic arm seems to be
totally aligned.}

The bright magnetic arms are located between the optical and gas
spiral arms, but they are not free of gas. The $\HI$ emission is
quite strong in the interarm regions of the inner galaxy, without
concentration to the magnetic arms (Fig.~\ref{HI}). Weak interarm
emission is seen in the maps of $7\mum$ dust emission
(Fig.~\ref{cm6i} left) and in H$\alpha$ (Fig.~\ref{th}). Deeper
images may reveal a relation between the magnetic arms and interarm
gas or dust features.

At \wave{18.0} and \wave{20.5}, where the signal-to-noise ratio is
highest, the bright magnetic arms observed at smaller wavelengths
turn out to be the brightest features of a system of at least five
magnetic arms which extend to at least 10\kpc\ radius
(Figs.~\ref{pol} bottom and Fig.~\ref{pol20}). The main northern
magnetic arm continues towards the east, 10\kpc\ from the centre.
Two more magnetic arms can be identified north and north-west of the
main northern magnetic arm. Both outer magnetic arms are located in
regions with very low density of $\HI$ gas (Fig.~\ref{HI}), one at
the outer edge of an $\HI$ arm at 3\farcm5 (5.5\kpc) north of the
centre, the other one at the inner edge of the next arm at 6\arcmin\
(9.5\kpc) north-west of the centre. South-east of the main southern
magnetic arm, another magnetic arm is seen which is broad and may
consist of several individual features. More magnetic arms could
exist in the outer south-western quadrant, but strong Faraday
depolarization prevents their detection at \wave{18.0} and
\wave{20.5}. The two bright magnetic arms can be interpreted as a
superposition of two azimuthal dynamo modes ($m=0$ and $m=2$), but
five magnetic arms or more need higher modes with $m\ge3$ (see
discussion in Sect.~\ref{sect:arms}).

The maps of polarized intensities at short wavelengths
(Fig.~\ref{pol} top) and those at longer wavelengths (Fig.~\ref{pol}
bottom) differ in the south-western quadrant where Faraday
depolarization $DP$ is strong at \wwave{18.0}{20.5}, as already
known from previous observations at lower resolution (Beck\
\cite{B91}). The origin of this $DP$ asymmetry is discussed in
Sect.~\ref{sect:depol}. The Effelsberg \wave{6.2} and \wave{11.2}
polarization maps (Figs.~\ref{pol6} and \ref{pol11}) do not show
this hole. This indicates that Faraday depolarization is not yet
strong at \wave{11.2}.

The orientations of the field vectors (Fig.~\ref{pol} right) follow
well the orientation of the spiral structure seen in polarized
intensity. A similar result has been found for M~51 where the
large-scale spiral shock front is able to align the magnetic field
(Patrikeev et al.\ \cite{P06}). However, spiral shocks cannot
explain the field alignment along the magnetic arms of NGC~6946 (see
Sect.~\ref{sect:arms}).

\subsection{Field regularity}
\label{sect:reg}

$p_n$ is a measure of the ratio $q$ of the field strength of the
regular field in the sky plane and the random field, the
\emph{degree of regularity}\ of the field,
$q=B_\mathrm{reg}/B_\mathrm{ran}$. As some fraction of the field
observed in polarization can be an anisotropic turbulent field (see
footnote~(3)), $q$ as derived from $p_n$ is an upper limit. For
constant cosmic-ray energy density and an isotropic random field
(Burn\ \cite{B66} and eq.~(2) in Beck et al.\ \cite{B+03}):

\begin{eqnarray}
p_n/p_0 & \, = & \, q^2 \, / \, (q^2+{2\over3}) \\
q & \, = & \, \{ \, 2 \, (p_n/p_0) \, / \, [3 \, (1-(p_n/p_0))]
\,\}^{0.5} \, .
\end{eqnarray}

\noindent $p_0$ is the intrinsic degree of polarization;
$p_0=(3+3\alpha_n)/(5+3\alpha_n)$, where $\alpha_n$ is the
nonthermal spectral index. In case of equipartition between the
energy densities of magnetic field and cosmic rays (eq.~(21) in Beck
et al.\ \cite{B+03}, adding the missing term $q^{-2}$):

\begin{eqnarray}
p_n/p_0 & \, = & \, (q^2+{7\over3}) \,\, / \,\, (q^2+3+{10\over9} q^{-2})
\,\, \simeq \, \, q^2 \, / \, (q^2+{1\over2}) \\
q \, & \simeq  & \, \{ \, (p_n/p_0) \, / \, [2 \, (1-(p_n/p_0))]
\,\}^{0.5} \, ,
\end{eqnarray}

\noindent where the approximation is accurate to better than 5\% for
$q<0.1$, 2--5\% for $0.1\le q \le 1$ and better than 2\% for $q>1$.
For the same value of $p_n$, the field regularity $q$ in
case of equipartition is smaller by 0.87 times compared to the value
for constant cosmic-ray density. For small values of $q$ and in case
of equipartition, $p_n$ is larger by 1.33 times compared to constant
cosmic-ray density.

\begin{figure}[htbp]
\centerline{\includegraphics[bb = 62 128 528 623,width=0.495\textwidth,clip=]{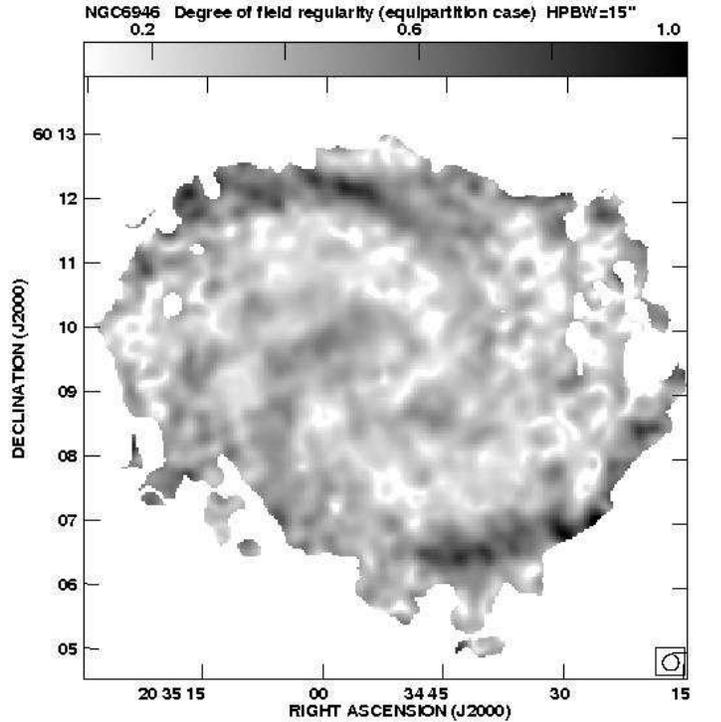}}
\caption{Degree of field regularity $q$, the ratio of the
regular field strength in the sky plane to the isotropic random
field strength, derived from the fractional polarization of the
nonthermal intensity of NGC~6946 at \wave{6.2} at 15\arcsec\
resolution. $q$ was computed only at points where the total
nonthermal intensity at \wave{6.2} exceeds 10 times the rms noise.}
\label{reg}
\end{figure}

Figure~\ref{reg} shows the degree of regularity $q$ of the
field, derived from the fractional polarization of the nonthermal
emission at \wave{6.2}, using eq.~(4) for the case of
equipartition. $q$ is generally below 1 over most of the galaxy,
but increases to about 1 in the outer parts of the two main magnetic
arms. In the local Milky Way, starlight polarization and radio
synchrotron data yield an average value of $q\simeq0.6$ (see Beck\
\cite{B01} for a review).

\subsection{Field compression in the inner spiral arms}
\label{sect:compr}

\begin{figure}[htbp]
\centerline{\includegraphics[bb = 62 164 528 601,width=0.495\textwidth,clip=]{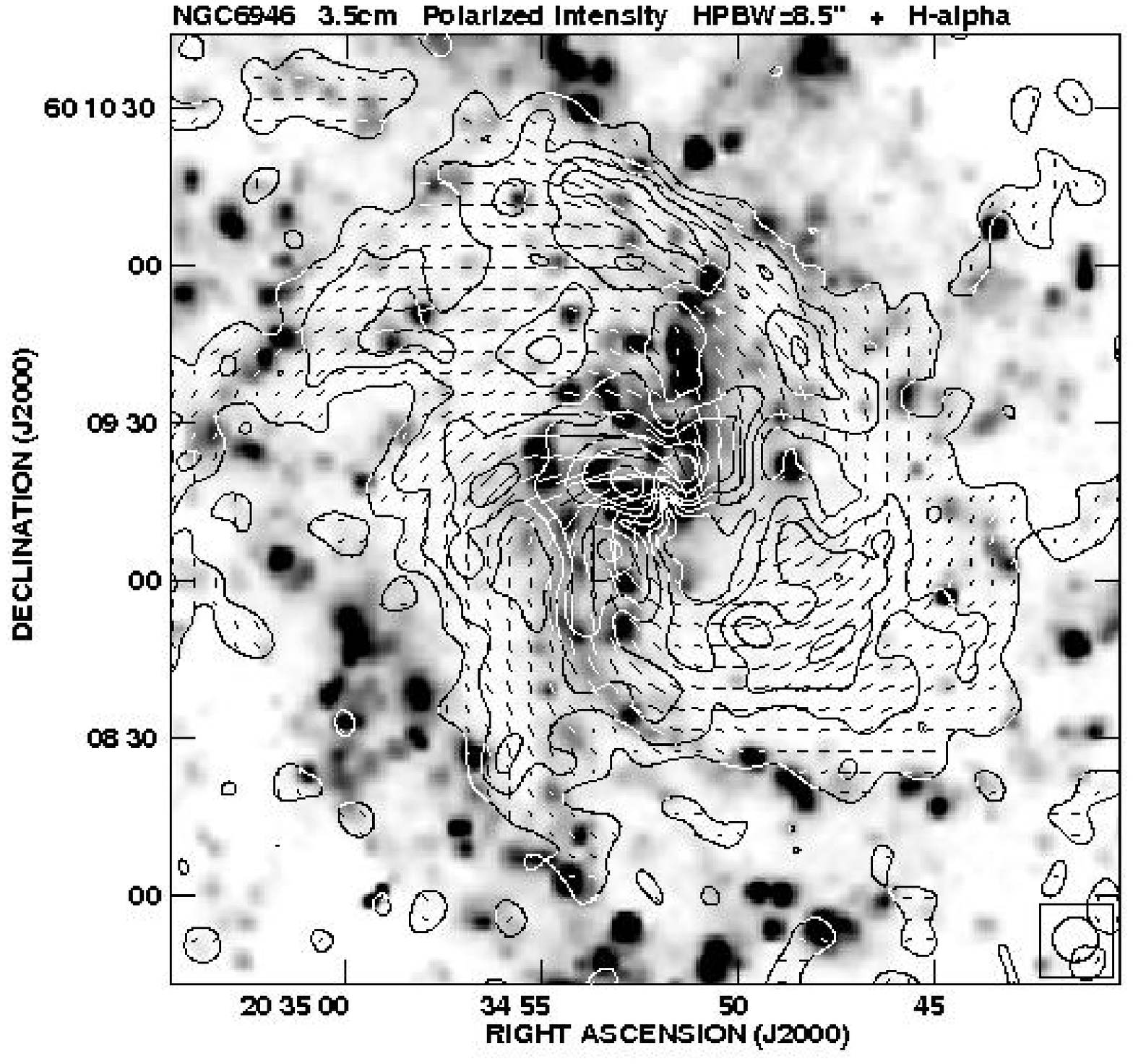}}
\caption{Polarized intensity contours and observed $B$-vectors
($E$+90\degr) of NGC~6946 at \wave{3.5} at 8\farcs5\ resolution,
observed with the VLA D-array. Faraday rotation is smaller than
5\degr\ at this wavelength. The contour intervals are $1, 2, 3, 4,
6, 8, 12 \times 20~\muJyb$. The vector length is proportional to
polarized intensity, 5\arcsec\ length corresponds to 100~$\muJyb$.
The background greyscale image shows the H$\alpha$ emission
(Ferguson et al.\ \cite{F98}). The position of the radio nucleus of
the galaxy is RA, DEC(2000) = $20^\mathrm{h}\ 34^\mathrm{m}\ 52\fs25$,
+60\degr\ 09\arcmin\ 14\farcs5.} \label{pol3}
\end{figure}

At the highest available resolution at \wave{3.5}, the spiral
pattern of the magnetic field continues well into the centre of
NGC~6946 (Fig.~\ref{pol3}). The polarized intensity is very diffuse
and fills almost the whole area, but avoiding the optical spiral
arms. High polarized intensities are found at the inner edge of the
inner spiral arm seen in Fig.~\ref{pol3} north of the centre, but
only along a small part of the southern inner optical arm. The
$B$-vectors are aligned along the northern optical arm, giving
evidence that the field is compressed or sheared, but
Faraday rotation is not enhanced in this region (see Fig.~\ref{rm}
left). This means that the field along the optical spiral arm is
{\em anisotropic turbulent}, with frequent (unresolved) reversals,
as expected from compression or shear of an isotropic
turbulent field. The compressed fields in the bars of NGC~1097 and
NGC~1365 (Beck et al.\ \cite{B+05}) and in the massive inner spiral
arms of M~51 (Fletcher et al.\ \cite{F07}) have similar properties.
{\em Polarized radio emission is an ideal tracer of field
compression or shear}. Polarization also nicely
traces interactions between galaxies or with the intergalactic
medium (Vollmer et al.\ \cite{V07}).

The contrast in polarized intensity between the northern inner arm
and the diffuse region inside of the arm is about 2. Assuming either
constant cosmic rays or energy equipartition between magnetic fields
and cosmic rays, this corresponds to a mild compression of the
magnetic field by $1.4\times$ and $1.2\times$, respectively.

The $B$-vectors in Fig.~\ref{pol3} are not sharply deflected in the
compression region, as expected for a shock front, but turn smoothly
from the upstream region (east of the northern arm) into the
compression region. A similar observation in barred galaxies lead to
the conclusion that the regular field is not coupled to the cold,
molecular gas but to the warm, diffuse gas (Beck et al.\ \cite{B05}).

The hole in polarized intensity south-west of the nucleus
(Fig.~\ref{pol3}) is the result of beam depolarization, as the field
is strongly twisted within the telescope beam. The offset with
respect to the galaxy nucleus could be due to asymmetry of the field
pattern, possibly due to shearing gas flows near the inner stellar bar
which is also observed in CO lines
(Schinnerer et al.\ \cite{S06}). The generation of strong fields with a
large-scale spiral pattern is typical for late-type barred galaxies
(Beck et al.\ \cite{B+02}, \cite{B05}).

\subsection{Faraday rotation}
\label{sect:rm}

\begin{figure*}[htbp]
\includegraphics[bb = 62 135 528 637,width=0.475\textwidth,clip=]{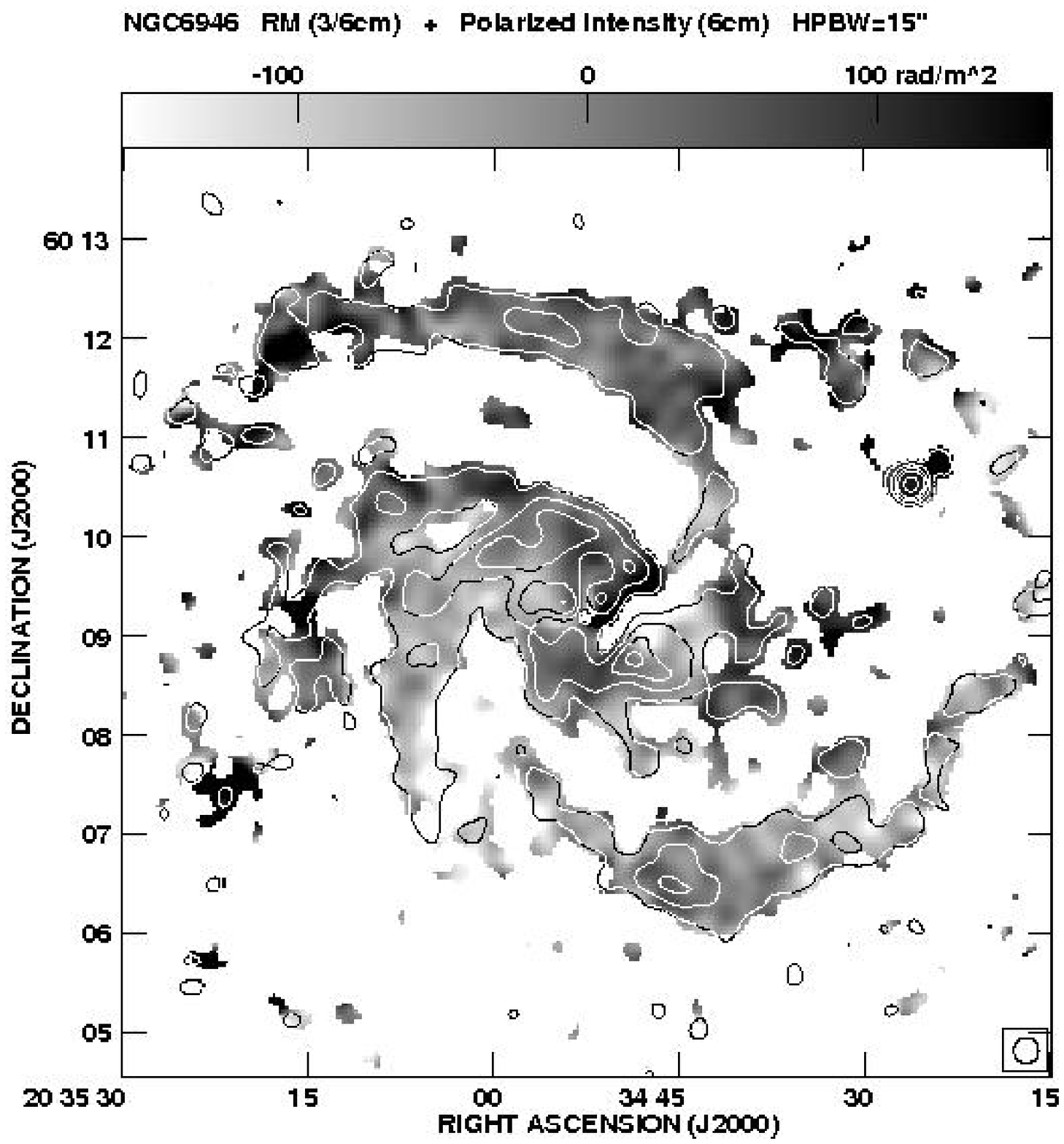}
\hfill
\includegraphics[bb = 62 135 528 637,width=0.475\textwidth,clip=]{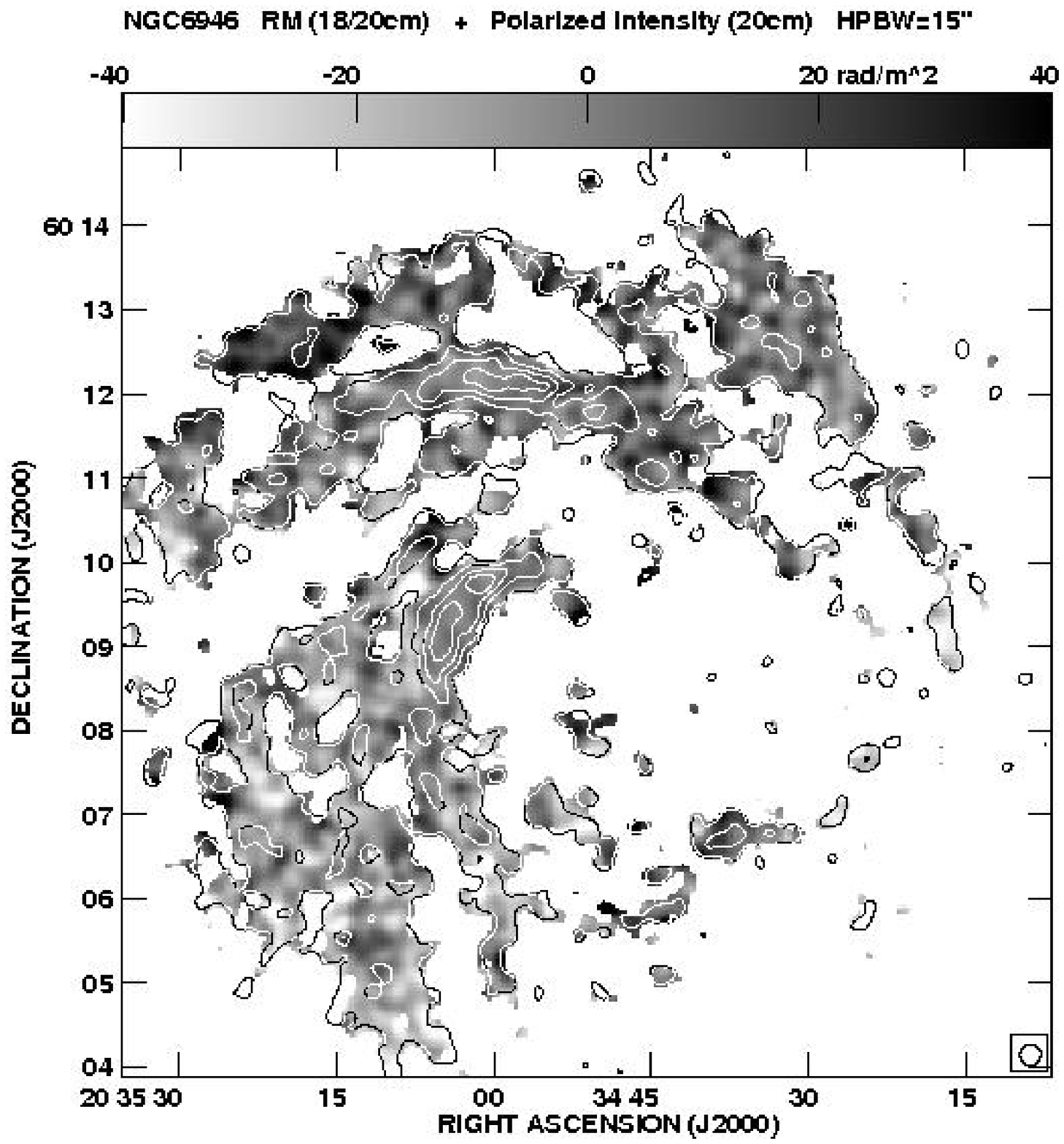}
\caption{ {\it Left:\/} Intrinsic Faraday rotation measures $RM_i$
between \wave{3.5} and \wave{6.2} at 15\arcsec\ resolution at points
where the polarized intensities at both wavelengths exceeds 4 times
the rms noise (taken from that in $Q$ and $U$). The contours show
the polarized intensity at \wave{6.2}. The contour intervals are $1,
2, 3, 4, 6 \times 80~\muJyb$. The foreground $RM_\mathrm{fg}$ from our
Galaxy of $40\radm$ has been subtracted. {\it Right:\/} Intrinsic
Faraday rotation measures $RM_i$ between \wave{18.0} and \wave{20.5}
at 15\arcsec\ resolution at points where the polarized intensities
at both wavelengths exceeds 4 times the rms noise. The contours show
the polarized intensity at \wave{20.5}. The contour intervals are
$1, 2, 3, 4 \times 60~\muJyb$. The foreground $RM_\mathrm{fg}$ from our
Galaxy of $40\radm$ has been subtracted.} \label{rm}
\end{figure*}

The polarization angles at \wave{3.5} and \wave{6.2} and those at
\wave{18.0} and \wave{20.5} were combined to maps of Faraday
rotation measures $RM$. Clipping below 4 times the rms noise in
polarized intensity restricts the maximum $RM$ error to $\pm69\radm$
and $\pm19\radm$, respectively. The average value in both maps of
$+40\pm2\radm$ has been adopted as the foreground $RM_\mathrm{fg}$ from
our Galaxy, consistent with previous results (Beck\ \cite{B91}, Ehle \&
Beck\ \cite{EB93}), and has been subtracted to obtain Faraday
rotation measures $RM_i$ intrinsic to NGC~6946 (Fig.~\ref{rm}). The
ambiguity of $RM$ values between \wave{3.5} and \wave{6.2}
of $n$ times $\pm 1232\radm$ due to the $\pm\pi n$ ambiguity
of polarization vectors is large and can be excluded.

$RM$ errors are large due to the small wavelength difference. Noise
in the maps of polarized intensity accounts for $RM$ fluctuations of
$\Delta RM = (\sqrt{2}\,S\,(\lambda_2^2-\lambda_1^2)\,)^{-1}$ where
$S$ is the signal-to-noise ratio of the polarized intensities at
$\lambda_1$ and $\lambda_2$. The rms noise in $PI$ is taken to be
the mean rms noise in $Q$ and $U$. $\Delta RM$ between \wave{3.5}
and \wave{6.2} due to noise is expected to increase from $12\radm$
in the brightest inner parts of the northern magnetic arm
($S\simeq23$), to $46\radm$ over the whole northern arm
($S\simeq6$), and to $69\radm$ at the cutoff limit in Fig.~\ref{rm}
of $S=4$. In the brightest part of the northern magnetic arm the
observed dispersion in $RM$ (Fig.~\ref{rm} left) is $25\radm$, so
that the contribution by random fields is $\simeq10\radm$, while the
dispersion of $\simeq50\radm$ observed over the whole arm is
dominated by fluctuations due to noise. A similar conclusion is
valid for the fluctuations in the $RM$ map between \wave{18} and
\wave{20} (Fig.~\ref{rm} right).

$RM_i$ between \wave{3.5} and \wave{6.2} (Fig.~\ref{rm} left) shows
the well-known north-south asymmetry with predominantly positive
values in the northern magnetic arm and negative ones in the
southern magnetic arm. The large-scale variation of $RM$ with
azimuthal angle in the galaxy plane is sinusoidal (Ehle \& Beck\
\cite{EB93}). This has been interpreted as a superposition of the
$m=0$ and $m=2$ dynamo modes (Rohde et al.\ \cite{R99}). \emph{No
other large-scale field reversals}\ are observed.

The $RM$ amplitude of $\pm 100\radm$ is provided by a regular
field of about $8\muG$ strength (Sect.~\ref{sect:energies}) and
diffuse ionized gas with an electron density of $\langle
n_\mathrm{e} \rangle \simeq 0.03\ccm$ and a pathlength of about
$1\kpc$.

In the Milky Way, Faraday rotation data from pulsars indicate
multiple large-scale field reversals (Han et al.\ \cite{H+06}).
Nothing similar has been observed in any external galaxy so far.
Although Faraday rotation data in galaxies like NGC~6946 are
line-of-sight integrals, large-scale field reversals would be
evident from the existing data. Furthermore, the general
concentration of coherent fields in interarm regions, like in
many spiral galaxies, was not found in the Milky Way. Either
our Galaxy is special, or the data are not yet sufficient to trace
the large-scale Galactic field.

$RM_i$ between \wave{18.0} and \wave{20.5} (Fig.~\ref{rm} right) is
on average 5 times smaller than $RM_i$ between \wave{3.5} and
\wave{6.2}, which is the effect of Faraday dispersion in the disk
(Sect.~\ref{sect:dp}). In the south-west quadrant no data are
available due to the weak polarized emission at \wave{18.0} and
\wave{20.5}. At a resolution of 72\arcsec\ $RM_i$ drops to about
$-50\radm$ in this region (Beck\ \cite{B91}).

A close pair of bright background sources (source A: RA, DEC(2000) =
$20^\mathrm{h}\ 34^\mathrm{m}\ 24\fs2$, +60\degr\ 10\arcmin\ 41\arcsec,
source B: $20^\mathrm{h}\ 34^\mathrm{m}\ 26\fs1$, +60\degr\ 10\arcmin\
32\arcsec) has fractional polarizations of 2\% and 12\% at both
\wave{3.5} and \wave{6.2}. Both sources are
completely depolarized at \wave{20}. At \wave{18} only source B is
still visible ($p\simeq3\%$). Their $RM$ between \wave{3.5} and
\wave{6.2} is $252\pm20\radm$ and $171\pm2\radm$. As polarized waves
from background sources travel through the whole disk, their $RM_0$
(after subtraction of $RM_\mathrm{fg}$) should be about twice that of
the internal emission $RM_i$. While the high $RM_0$ of source A is
probably intrinsic, $RM_0\simeq131\radm$ of source B corresponds to
$RM_i\simeq65\radm$ of the diffuse emission, consistent with the
$RM_i$ of the diffuse emission of NGC~6946 observed in this region
(Fig.~\ref{rm} left).

Another pair of bright, polarized background sources, barely
separated, is visible at the southern edge of the maps at \wave{18}
and \wave{20} (Fig.~\ref{HI}), located at RA, DEC(2000) =
$20^\mathrm{h}\ 35^\mathrm{m}\ 19\fs2$, +60\degr\ 02\arcmin\ 08\arcsec,
at 12\kpc\ distance from the centre of NGC~6946. The total flux
densities are $860\pm50\muJy$ at \wave{18} and $1070\pm40\muJy$ at
\wave{20}. The rotation measure, corrected for the average foreground
of $RM_\mathrm{fg}=40\pm2\radm$, is $RM_0 = -13\pm4\radm$ and may
indicate that regular fields and ionized gas in NGC~6946 extend to at
least 12\kpc\ radius. However, a variation of $RM_\mathrm{fg}$ cannot be
excluded. Ehle \& Beck (\cite{EB93}) determined
$RM_\mathrm{fg}=30\pm4\radm$ for the radius range
9.5--14\kpc\ which means no significant internal rotation measure
($RM_0 = -3\pm6\radm$) towards the background sources. Observations
of more polarized sources with higher sensitivity are required.

\subsection{Structure function of $RM$}
\label{sect:sf}

Faraday rotation measures are signatures of structures of coherent
regular magnetic fields in an astrophysical plasma with a
spectrum of scales which can be characterized in various ways.
Wavelet spectra are an excellent tool to analyze intensity maps
(Frick et al.\ \cite{F01}), but cannot be applied to $RM$ data.
Analysis of $RM$ data of galaxy clusters with the autocorrelation
function revealed power spectra of the magnetic field approximately
of Kolmogorov type (Vogt \& En{\ss}lin\ \cite{V03}). In our Galaxy,
on scales of 0.5--5\degr\ ($\simeq$10--100\pc\ at 1~kpc distance),
Haverkorn et al. (\cite{H06}) found slopes between 0.4 and 0.8 in
interarm regions, but very flat $SFs$ in optical spiral arms. The
resolution obtained in NGC~6946 allows to study only scales larger
than about $400\pc$.

\begin{figure}[htbp]
\centerline{\includegraphics[bb = 76 349 500 666,width=0.475\textwidth,clip=]{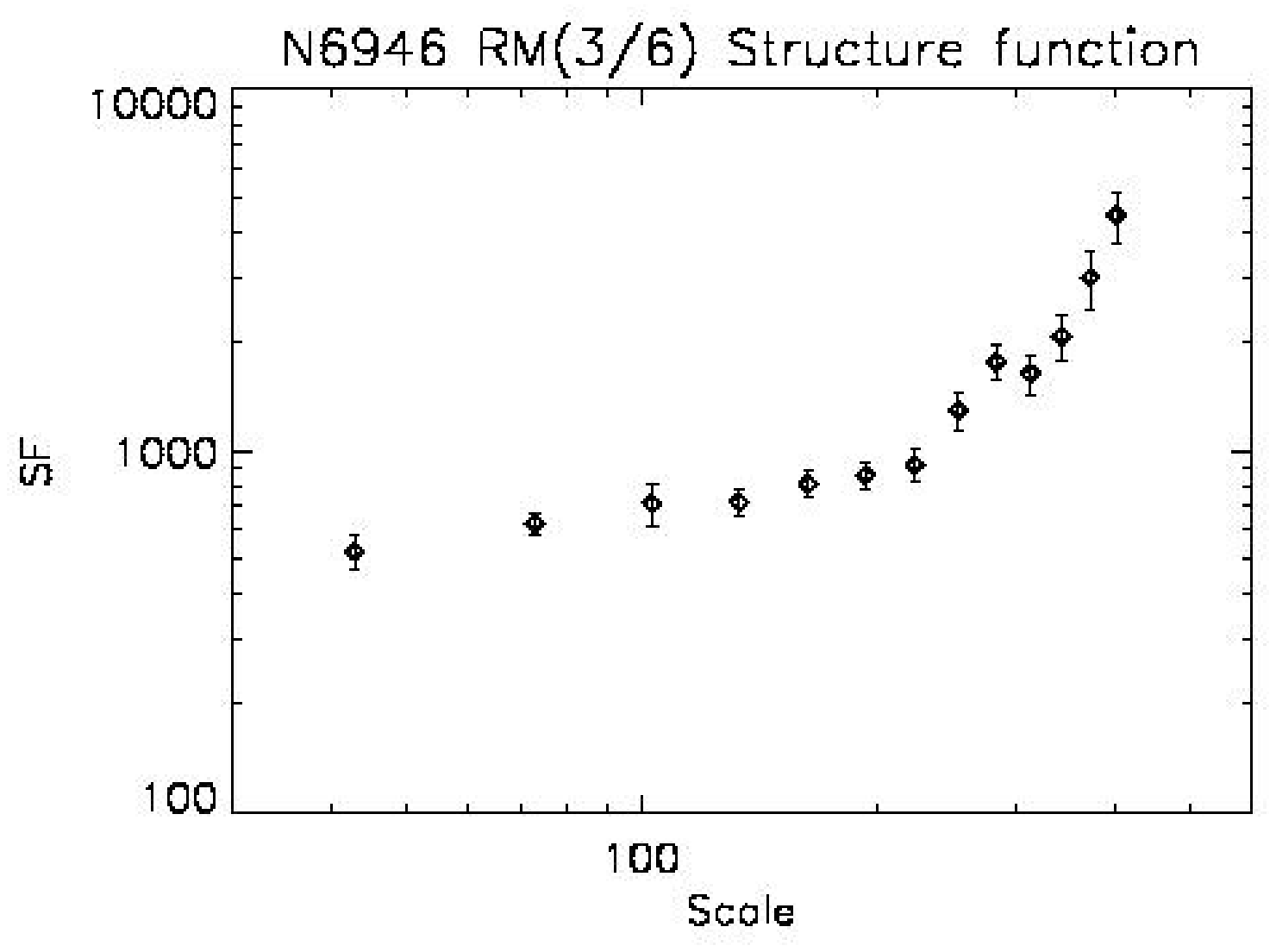}}
\hfill
\centerline{\includegraphics[bb = 98 349 500 666,width=0.475\textwidth,clip=]{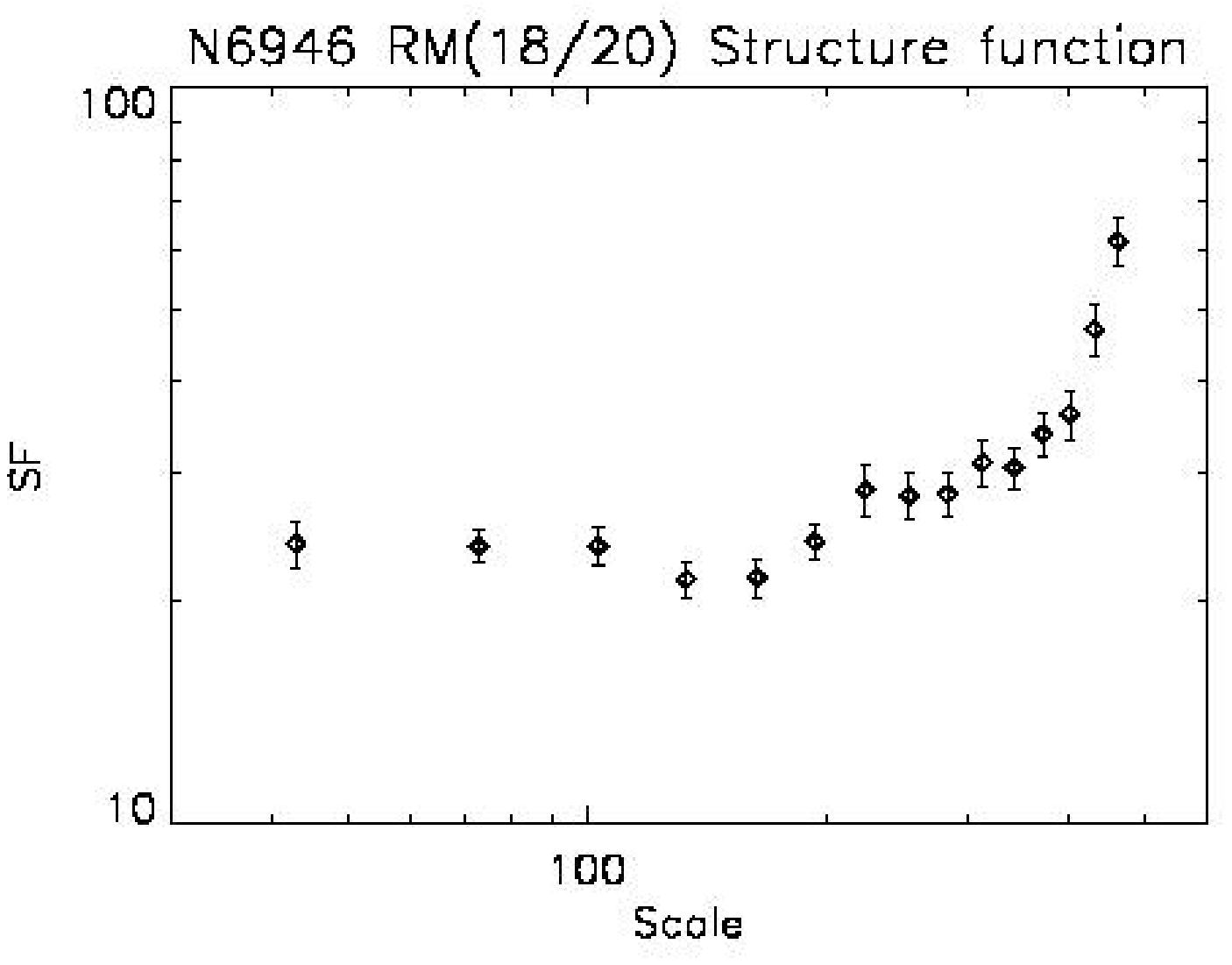}}
\caption{Structure function (in logarithmic presentation) of the
intrinsic Faraday rotation measures $RM_i$ between \wave{3.5} and
\wave{6.2} (top) and between \wave{18} and \wave{20} (bottom) at
15\arcsec\ resolution. The scale is measured in arcseconds.}
\label{sf}
\end{figure}

Figure~\ref{sf} shows the structure function $SF$ of the intrinsic
$RM_i$ values in NGC~6946 at the two wavelength pairs. The
errors in $SF$ were computed from the standard deviations. The $SF$
of $RM_i$ between \wave{3.5} and \wave{6.2} is flat (slope
$\simeq0.3$) until 4\arcmin\ (6\kpc) and then steepens (slopes of
$\simeq$~2--4). The $SF$ of $RM_i$ between \wave{18} and \wave{20} is
very flat (slope $\simeq0.0$) until 3\arcmin\ (5\kpc) and then
mildly steepens (slope $\simeq0.3$). Hence, Faraday effects shift
spectral energy from large to small scales.

The $RM_i$ structures seen in NGC~6946 are correlated on scales of
$\ge6\kpc$ which is due to the magnetic arms hosting coherent
regular fields. On scales of 1--6\kpc\ the $RM_i$ distribution is
mostly chaotic, without indications of systematic structures in the
coherent magnetic fields, like Parker loops. Coherent fields on kpc
scales also exist in the Milky Way, measured from pulsar rotation
measures (Han et al.\ \cite{H+06}) and from Zeeman splitting of OH
maser lines (Fish et al.\ \cite{FR03}).

\subsection{Faraday depolarization}
\label{sect:dp}

The new polarization maps at \wave{18.0} and \wave{20.5} at
15\arcsec\ resolution (Fig.~\ref{pol} bottom) do not show more
polarized emission in the SW quadrant than the previous data at
lower resolution (Beck\ \cite{B91}). Therefore 2-D depolarization
mechanisms acting in the sky plane, like Faraday dispersion in a
foreground screen or $RM$ gradients, are less important in NGC~6946.
3-D Faraday depolarization mechanism within the emitting volume,
i.e. differential Faraday rotation or Faraday dispersion, dominate.
In contrast to NGC~6946, the systematic pattern of depolarization in
M~31 shows that most of the depolarization in that galaxy is due to
foreground $RM$ gradients. Compared to NGC~6946, $RMs$ in M~31 are
larger because the coherent regular field is stronger and the
galaxy's inclination is higher (Fletcher et al.\ \cite{F04}).

\begin{figure*}[htbp]
\includegraphics[bb = 62 135 528 637,width=0.475\textwidth,clip=]{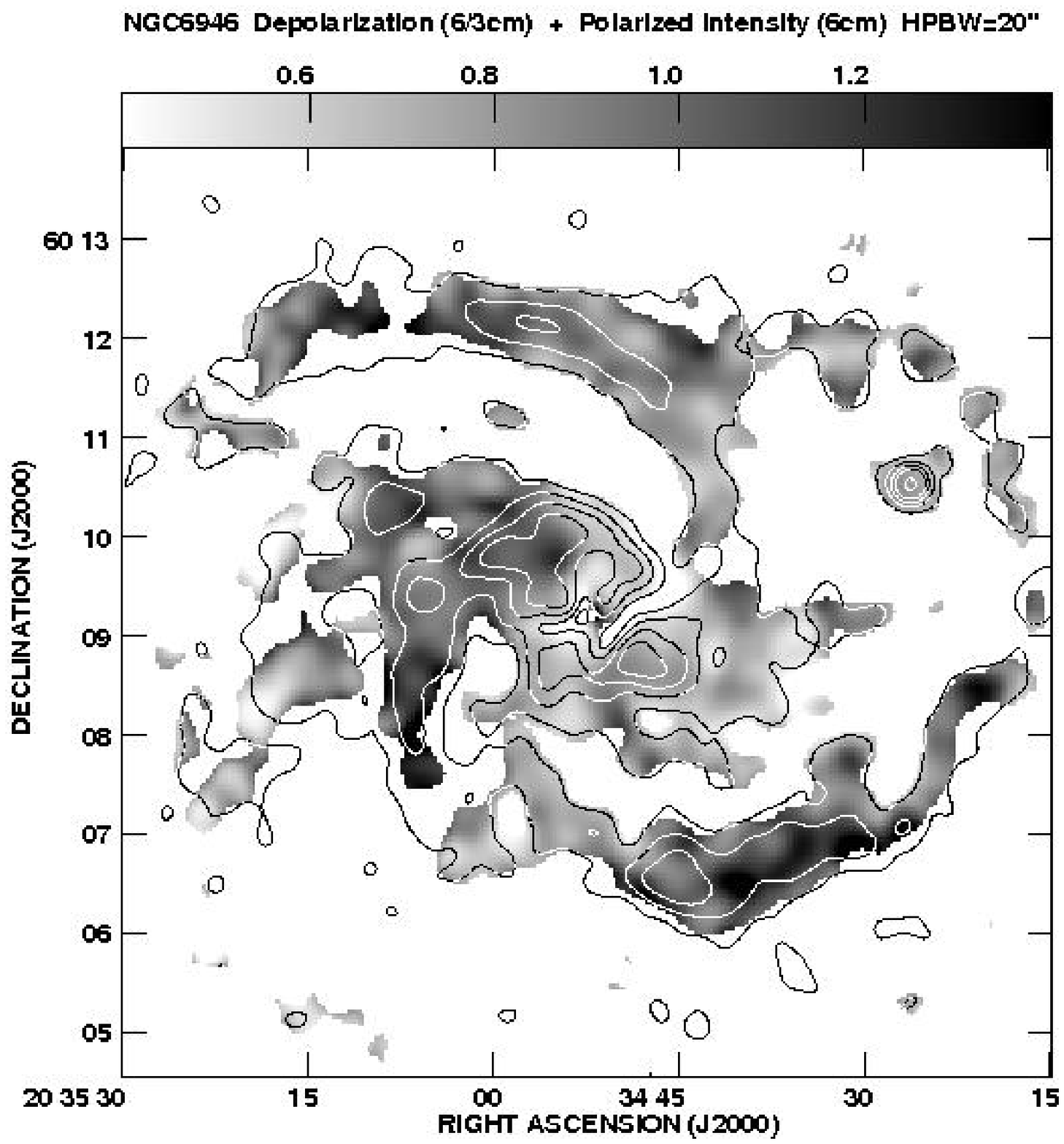}
\hfill
\includegraphics[bb = 62 135 538 637,width=0.475\textwidth,clip=]{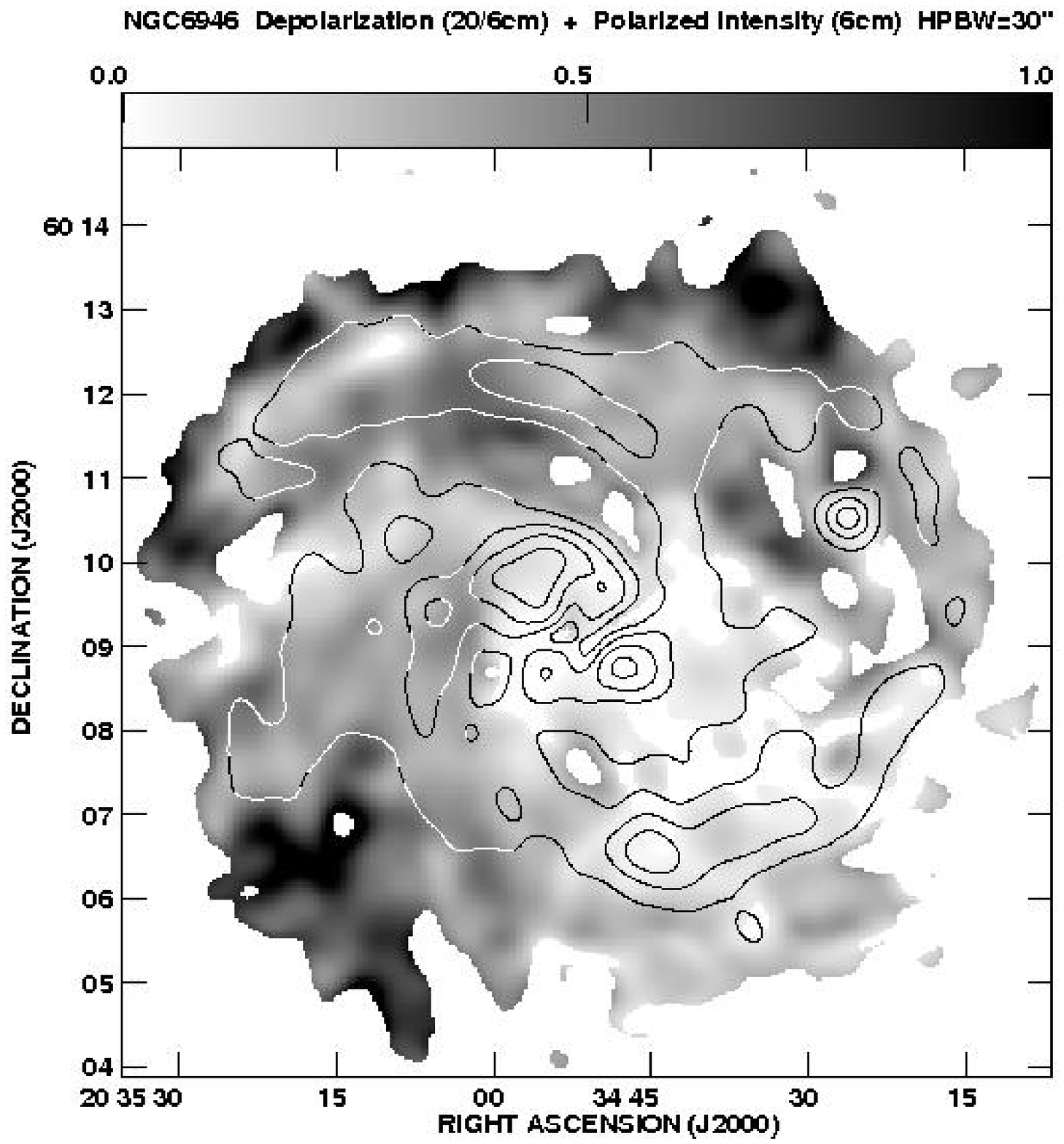}
\caption{ {\it Left:\/} Ratio between polarized intensities at
\wave{6.2} and \wave{3.5} at 20\arcsec\ resolution, corrected for
spectral index, as a measure of Faraday depolarization. The ratio
was computed only at points where the polarized intensities at both
wavelengths exceeds 4 times the rms noise (taken from that in $Q$
and $U$). The contours show the polarized intensity at \wave{6.2}.
The contour intervals are $1, 2, 3, 4, 6 \times 100~\muJyb$. {\it
Right:\/} Ratio between polarized intensities at \wave{20.5} and
\wave{6.2} at 30\arcsec\ resolution, corrected for spectral index,
as a measure of Faraday depolarization. The ratio was computed only
at points where the polarized intensities at both wavelengths
exceeds 4 times the rms noise. The contours show the polarized
intensity at \wave{6.2}. The contour intervals are $1, 2, 3, 4
\times 200~\muJyb$. } \label{dp}
\end{figure*}

Faraday depolarization is usually defined as the ratio $DP$ of the
degrees of polarization synchrotron emission at two wavelengths.
This requires subtraction of the thermal emission which is subject
to uncertainties (see Sect.~\ref{sect:thermal}). Instead, $DP$ was
computed in this paper by: $DP=(PI_1/PI_2) \times
(\nu_2/\nu_1)^{\alpha_n}$ where $\alpha_n=-1.0$ is the synchrotron
spectral index, assumed to be constant across the galaxy. Deviations
in $\alpha_n$ affect $DP$ less severely than the estimate of thermal
fraction of the radio emission.

$DP(6cm/3cm)$ (Fig.~\ref{dp} left) is around 1 in both main magnetic
arms and in the inner galaxy east of the centre. Smaller values
around 0.7, i.e. stronger depolarization, are found 1--3\arcmin\
south-west of the centre. $DP(20cm/6cm)$ (Fig.~\ref{dp} right) varies
over a much larger range. While $DP$ in the northern magnetic arm is
around 0.4, strong depolarization ($DP$ between 0.1 and 0.2) occurs
in the south-western quadrant.

Differential Faraday rotation within the emitting layer leads to
depolarization which varies with $\lambda$ as a $sin(x)/x$ function,
with $x=2 \, |RM_i| \, \lambda^2$ (Sokoloff et al.\ \cite{S98}). At
\wave{20.5} strong $DP$ is expected for $|RM_i| > 30\radm$, with
many lines of zero polarization (``canals'') along level lines with
$|RM_i|=n \, 37.5\radm$. Only two canals were identified which are
probably generated in the Galactic foreground
(Sect.~\ref{sect:rings}). Hence, differential Faraday rotation in
NGC~6946 cannot explain the large-scale asymmetry in $DP$
(Sect.~\ref{sect:depol}).

The intrinsic rotation measures between \wave{18.0} and \wave{20.5}
are a factor of about five smaller than those between \wave{3.5} and
\wave{6.2}. The galaxy is not transparent to polarized decimeter
radio waves, it is {\em Faraday thick}\ (Beck\ \cite{B91}; Ehle \&
Beck\ \cite{EB93}). Berkhuijsen et al. (\cite{B97}) successfully
modeled the similar situation in the spiral galaxy M~51 by assuming
Faraday dispersion, which depolarizes most of the disk, plus a
foreground layer in the upper disk or in the halo which have a lower
polarized intensity and rotation measure than the full disk.

Internal Faraday dispersion by turbulence in the magneto-ionic
interstellar medium along the line of sight is the probable source
of depolarization at long wavelengths. It is computed as (Sokoloff
et al.\ \cite{S+98}):

\begin{equation}
p\, = \, p_0 \, [1-\exp(-2S)\}]\, / \, (2S)\, ,
\end{equation}

\noindent where $S=\sigma_\mathrm{RM}^2 \lambda^4$. $\sigma_\mathrm{RM}$
is the dispersion in intrinsic rotation measure $RM_i$. (Note that a
factor of two was missing in Beck\ \cite{B91} and in Ehle \& Beck\
\cite{EB93}.) $DP=0.2$ at \wave{20.5} needs a dispersion of
$\sigma_\mathrm{RM}\simeq38\radm$.

Faraday dispersion is an effect of the turbulent interstellar medium
and can be written as $\sigma_\mathrm{RM}=0.81 n_\mathrm{e} B_\mathrm{r}
d (L f/d)^{0.5}$ where $n_\mathrm{e}$ is the thermal electron density of
the diffuse ionized gas (in $\ccm$), $B_\mathrm{r}$ the random field
strength (in $\muG$), $L$ the pathlength through the thermal gas (in
\pc), $d$ the turbulent scale (in \pc), and $f$ the filling factor of
the Faraday-rotating gas. Faraday rotation and depolarization occur in
the diffuse ionized medium with a large filling factor, while the
thermal optical and thermal radio emission is dominated by $\HII$
regions with a small filling factor (see discussion in Ehle \& Beck\
\cite{EB93}). Standard values of $n_\mathrm{e}=0.03\ccm$,
$B_\mathrm{r}=10\muG$, $L=1000\pc$, $d=50\pc$ and $f=0.5$ yield the
required $\sigma_\mathrm{RM}$.

In the south-western quadrant $DP$ decreases to $<0.2$
(Fig.~\ref{dp} right). If due to Faraday dispersion, an increase of
$n_\mathrm{e}$ or $B_\mathrm{r}$ is required, although star-formation
activity and total radio emission is lower in this quadrant. The $DP$
asymmetry is probably an effect of field geometry
(Sect.~\ref{sect:depol}).

\subsection{Depolarization rings}
\label{sect:rings}

\begin{figure*}[htbp]
\includegraphics[bb = 69 121 507 637,angle=270,width=0.475\textwidth,clip=]{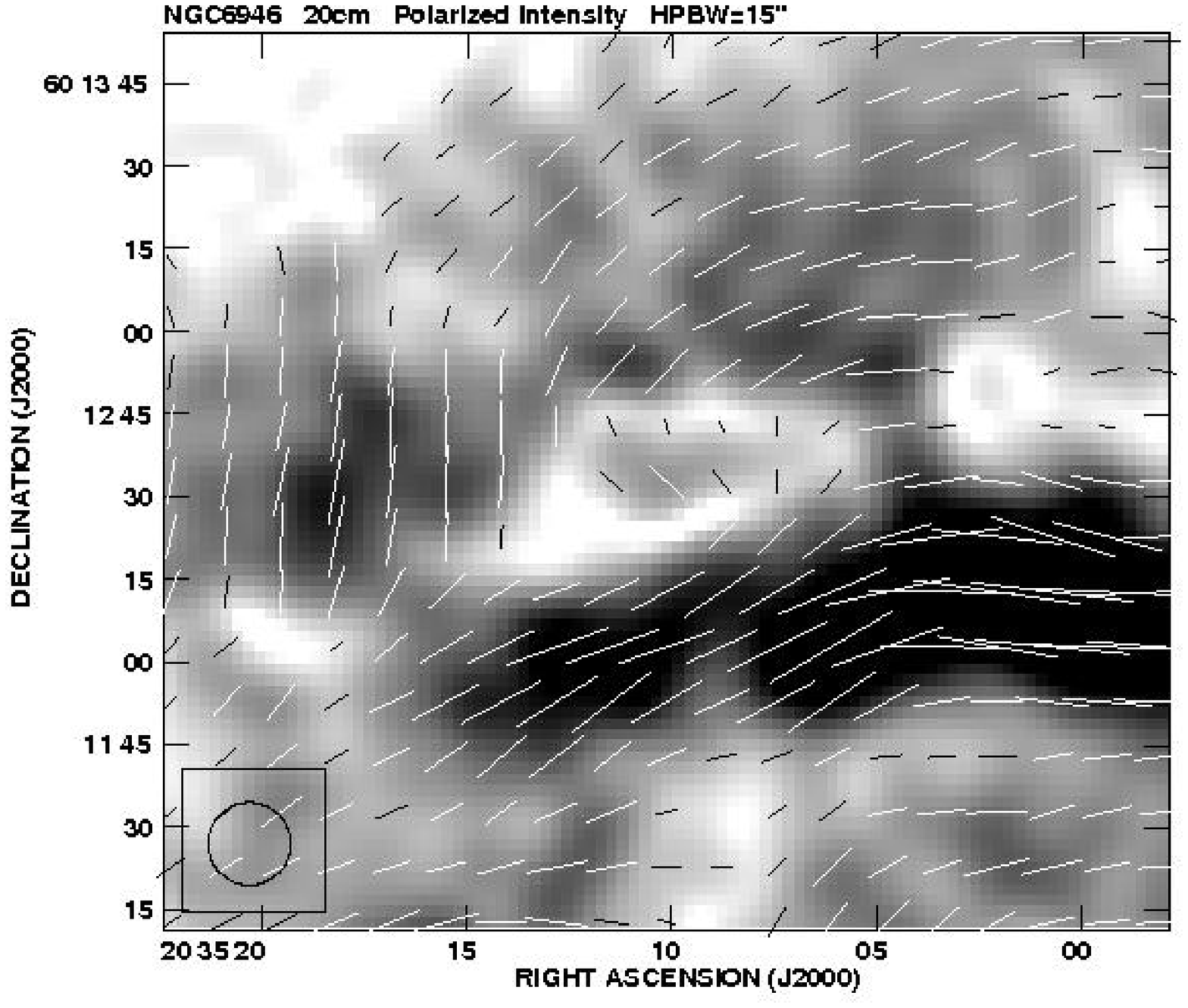}
\hfill
\includegraphics[bb = 62 106 507 652,angle=270,width=0.475\textwidth,clip=]{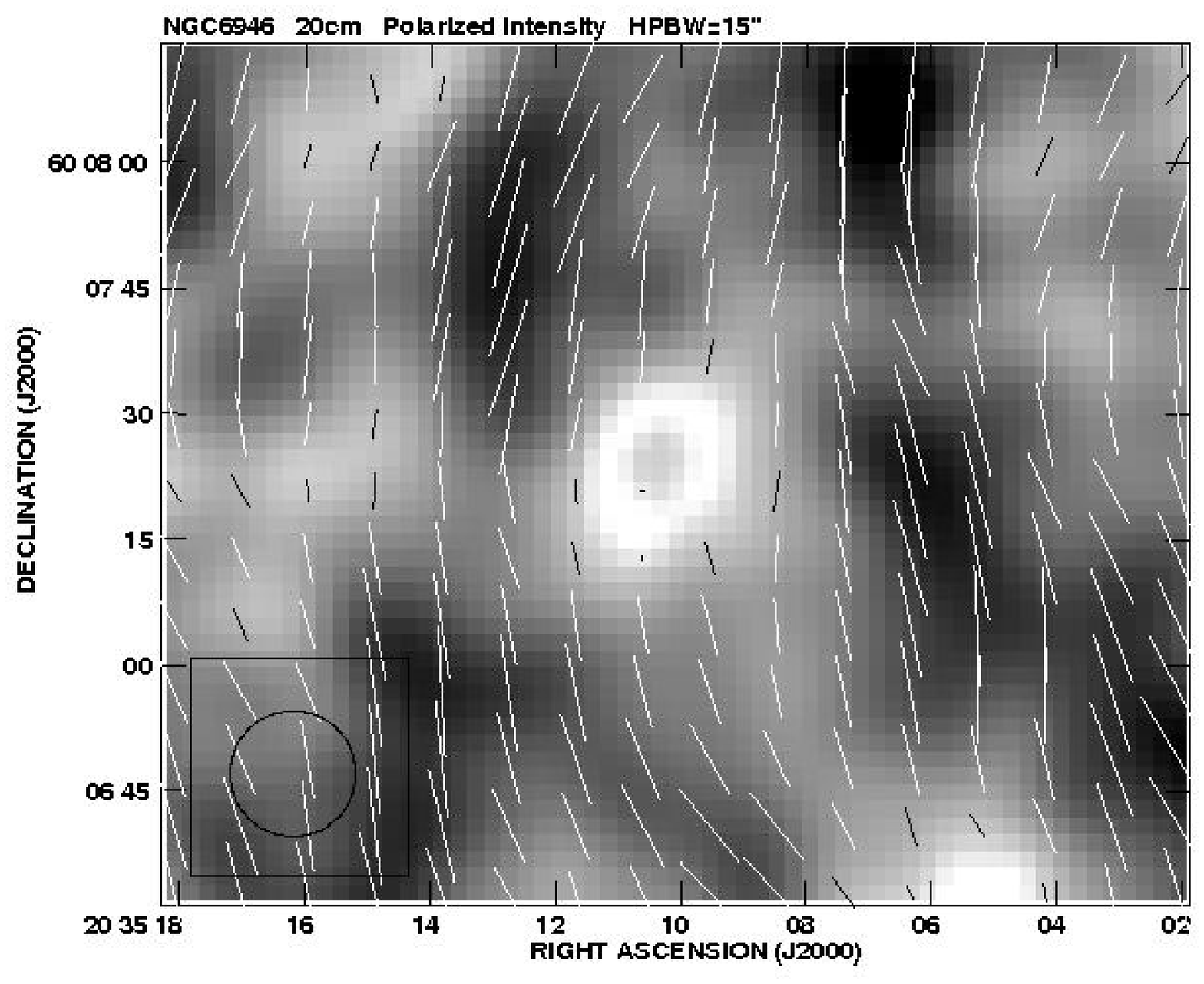}
\caption{Polarized intensity and observed $E$-vectors at \wave{20.5}
at 15\arcsec\ resolution in the regions showing ring-like
``canals''. } \label{canals}
\end{figure*}

The \wave{20.5} polarization map shows three narrow rings of zero
intensity (Fig.~\ref{canals}) with a 90\degr\ jump of the
polarization angle, resembling the \emph{canals}\ in the \wave{21}
and \wave{90} polarization maps of the Milky Way with resolutions of
more than an arcminute (Gaensler et al.\ \cite{G+01}; Uyan{\i}ker et
al.\ \cite{U+03}; Haverkorn et al.\ \cite{H04}) and in the \wave{20}
polarization map of M~31 with 45\arcsec\ resolution (Shukurov \&
Berkhuijsen\ \cite{SB03}). These Galactic canals have a much longer
angular extent than those in NGC~6946, most of them are not closed,
and the few closed ones are irregular in shape.

The half-power widths\footnote{Though structures in $Q$ or $U$
cannot be smaller than the beamsize, this is not true for maps of
polarized intensity. Canals are the result of a sign change in the
maps of Stokes $Q$ or $U$ or both (Fletcher \& Shukurov\
\cite{FS06}). A steep gradient in Stokes $Q$ or $U$ may lead to a
canal which is narrower than the beam.} of the rings in
Fig.~\ref{canals} are 4--6\arcsec. The ring diameters are 20\arcsec\
(ring at RA, DEC(2000) = $20^\mathrm{h}\ 35^\mathrm{m}\ 10^\mathrm{s}$,
+60\degr\ 12\arcmin\ 35\arcsec) and 12\arcsec\ (rings at RA, DEC(2000) =
$20^\mathrm{h}\ 35^\mathrm{m} 02^\mathrm{s}$,
+60\degr\ 12\arcmin\ 50\arcsec\ and at $20^\mathrm{h}\ 35^\mathrm{m}\
10^\mathrm{s}$, +60\degr\ 07\arcmin\ 25\arcsec), corresponding to
530\pc\ and 320\pc\ if generated in NGC~6946, or to $0.1\pc\times D$
and $0.06\pc\times D$ at a distance $D$ (in \kpc) in the Galactic
foreground.

Depolarization canals or rings can be generated in the Milky Way by
a superposition of diffuse background emission which is rotated in a
\emph{Faraday screen}\ with unrotated foreground emission (Sun et
al.\ \cite{S+07}). As our VLA synthesis observations were not
sensitive to the diffuse Galactic foreground emission, this
explanation cannot be applied.

The foreground $RM_\mathrm{fg}$ in the Milky Way with a mean value of
$40\radm$ can also generate a depolarization canal or ring if a jump
or gradient of $\Delta RM_\mathrm{fg}=(\pm n + {1\over2}) \,
\pi/\lambda^2$ occurs on scales smaller than the beamsize (Fletcher \&
Shukurov\ \cite{FS06}). The rings of NGC~6946 could be signatures of
small-scale structures in the magnetized Galactic interstellar
medium with diameters of $d \simeq 0.06\pc \times D$, smaller than
those observed in Galactic surveys so far, in the size range of
Galactic planetary nebulae or compact $\HII$ regions. Two of the
rings have almost circular geometry. Assuming a spherical shape and
a constant regular field strength of $2\muG$ inside and outside the
object, a $RM$ jump of $38\radm$ requires a mean electron density of
$<n_\mathrm{e}> \simeq 400\ccm / D$ where $D$ is the distance in kpc.
With a 15\arcsec\ beam at 1.4~GHz, the thermal radio flux density per
beam is $\simeq 0.6\muJy\times EM$ where $EM$ is the emission measure.
With the numbers quoted above, $EM = <n_\mathrm{e}>^2 d / f \simeq 10^4$
cm$^{-6} \pc / D / f$, where $f$ is the filling factor, yielding a
large thermal flux density of several \mJyb, which is not detected
in Fig.~\ref{cm20i}, so that the rings cannot originate in
the Galactic foreground.

Shukurov \& Berkhuijsen (\cite{SB03}) and Fletcher \& Shukurov
(\cite{FS06}) discussed that the canals can also be generated by
internal differential Faraday rotation within the emitting medium.
Such canals appear at \wave{20.5} at level lines of $RM_i=\pm n \,
\pi/(2 \lambda^2)=\pm n \, 37.5\radm$ and at \wave{18.0} at $\pm n
\, 48.4\radm$ (Fletcher \& Shukurov\ \cite{FS06}). (Note that $RM_i$
is the observed internal rotation measure which is half of the
$RM_0$ through the whole source.) The intrinsic rotation measure
which is relevant for internal differential Faraday rotation is that
of the visible layer, i.e. the intrinsic rotation measure between
\wave{18.0} and \wave{20.5} (Fig.~\ref{rm} right). The $RM_i$ values
around the rings are smaller than the critical values for total
internal depolarization, so that a steep increase towards the rings
is required. Assuming a diameter $d \simeq 300\pc$ (e.g. a $\HII$
complex), a spherical shape and a constant regular field strength of
$3\muG$ inside and outside the object, a $RM$ jump of $38\radm$
requires a mean electron density of $<n_\mathrm{e}> \simeq 0.1\ccm$. The
corresponding emission measure of $EM \simeq 3$ cm$^{-6}\pc / f$
yields a thermal flux density of $\simeq 40 \muJyb$ for $f=0.05$,
only 2 times the rms noise at \wave{20.5} and hence not visible in
Fig.~\ref{cm20i}.

\section{Discussion}

\subsection{The origin of the magnetic spiral arms}
\label{sect:arms}

The two bright inner magnetic spiral arms of NGC~6946 are \emph{not}
the effect of reduced depolarization in the regions between the
optical spiral arms. Firstly, wavelength-dependent Faraday
depolarization is not weaker (indicated by larger $DP$ values) in
the magnetic arms (Fig.~\ref{dp} right). Secondly, reduced
wavelength-independent depolarization (less tangled fields or less
turbulent fields) can be excluded by the fact that the magnetic arms
are seen also in total intensity (Figs.~\ref{cm6i} and \ref{cm20i}
right). The intensity enhancement in the magnetic arms above the
background is about the same in polarized and in total intensity.
Hence, the magnetic arms are real features where the strengths of
the regular {\em and}\ of the total field are enhanced. After
subtraction of the smooth, unpolarized disk emission, the degree of
polarization is very high in the magnetic arms, reaching the maximum
possible value of 75\% in the northern magnetic arm, so that the
magnetic field is totally aligned there.

Magnetic arms exist in several other gas-rich spiral galaxies of
Hubble type Sc, for example in IC~342 (Krause\ \cite{K93}; Beck\
\cite{B05}), NGC~2997 (Han et al.\ \cite{H99}) and M~83 (Beck\
\cite{B05}), but are less prominent than in the classical case
NGC~6946. In galaxies with strong density waves, like M~51, bright
ridges at the inner edges of the spiral arms dominate the maps of
polarized intensity (Patrikeev et al.\ \cite{P06}), but the diffuse
polarized emission in the interarm regions may have a similar origin
as the magnetic arms. Sb-type galaxies like M~31 and M~81 and
flocculent galaxies like M~33 do not host magnetic arms (see Beck\
\cite{B05} for a review). Hence, the phenomenon of magnetic arms
seems to occur preferably in galaxies with a large mass of cold gas,
a high star-formation rate and strong magnetic fields.

The analysis technique with 1-D wavelet functions by Frick et al.
(\cite{F00}) is a tool to measure the pitch angle of spiral arm
structures. Application to the NGC~6946 data indicated that the
magnetic arms seen in polarization are phase-shifted ``images'' of
the gas/optical arms, receding in the sense of galactic rotation,
and hence could be the result of interaction between the gaseous
density wave and the magnetic dynamo wave. The structure of the
magnetic arms is more regular than that of the gas arms. Gas inside
the corotation radius of $\simeq 10\kpc$ (Rohde et al.\ \cite{R99})
rotates faster than the spiral pattern and crosses the magnetic arm
before entering the gas arm, so that the magnetic arm is generated
first. \emph{The properties of the gas arm could be determined by
the magnetic arm}.

{\em Slow 2-D MHD waves}\ in a galactic disk provide the required
phase shift between the amplitudes of the gas and the magnetic field
and also explain the alignment of the pitch angles of the magnetic
field (Fan \& Lou\ \cite{F96}, \cite{F97}; Lou \& Fan\ \cite{L98}).
However, magnetic arms are observed only in the differentially
rotating part of NGC~6946 while in the model slow MHD waves exist
only in the rigidly rotating part. Furthermore, slow MHD waves are
unstable in three dimensions as they evolve into the modes of the
Parker instabilities (Foglizzo\ \& Tagger\ \cite{FT95};
Shukurov\ \cite{S05}).

Linear models of the {\em turbulent galactic dynamo}\ are in
principle able to explain the two bright magnetic arms if dynamo
action in the interarm regions is enhanced by increased magnetic
diffusivity or a larger correlation length (Moss\ \cite{M98};
Shukurov\ \cite{S98}; Rohde et al.\ \cite{R99}). The $m=0$ azimuthal
dynamo mode generates an axisymmetric spiral field with constant
field direction, the $m=2$ azimuthal mode four magnetic spiral arms
with alternating field directions. If the $m=0$ and $m=2$ azimuthal
modes are superimposed with the same amplitudes, two arms remain
which have the same direction of the radial field component with
respect to the galaxy's centre, in agreement with the large-scale
$RM$ distribution observed in NGC~6946 (Sect.~\ref{sect:rm}).
However, the dynamo model predicts that the phase shift between the
magnetic arm and the optical arm varies with galacto-centric radius
and vanishes at the corotation radius, while the observed phase
shift is almost constant with radius (Frick et al.\ \cite{F00}).

In the interpretation of the magnetic arms as dynamo modes, the
pitch angle of the field vector depends on the dynamo numbers which
reflect the physical conditions required for dynamo excitation
(Shukurov\ \cite{S00}). If these conditions vary locally, the pitch
angle will also vary. Rohde et al. (\cite{R99}) showed that the
average absolute field pitch angle in NGC~6946 is indeed smaller by
about 10\degr\ in the magnetic arms than in the nearby optical arms.

\begin{figure}[htbp]
\centerline{\includegraphics[bb = 62 135 550 637,width=0.495\textwidth,clip=]{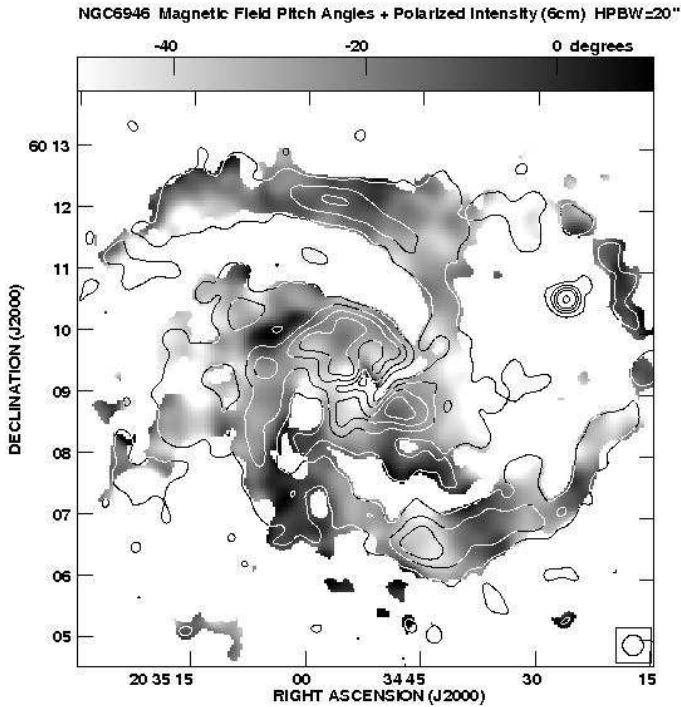}}
\caption{Pitch angles of the magnetic field vectors at 20\arcsec\
resolution and contours of polarized intensity at \wave{6.2}. }
\label{pitch}
\end{figure}

To quantify the regularity of the spiral field pattern in NGC~6946,
field pitch angles were computed from the observed $B$-vectors. At
each point of the map with sufficiently high polarized intensity,
the position angle of the circumferential vector in the plane of
galaxy, projected into the sky plane, was subtracted from the
position angle of the Faraday-corrected $B$-vectors (Fig.~\ref{pol}
top right). The plane of the galaxy was assumed to be inclined by
38\degr\ to the sky plane. The major axis of the galaxy plane
projected to the sky plane was assumed to have a position angle of
60\degr\ (Carignan et al.\ \cite{C90}). Figure~\ref{pitch} shows the
resulting map of the {\em pitch angles of the magnetic field
vectors}\ in the sky plane. The average pitch angle of the field
vectors in the northern magnetic arm is $-20\degr\pm10\degr$, with a
tendency of decreasing absolute values from the inner to the outer
part. The rms variations are consistent with those expected from rms
noise in the original maps. The field in the southern main magnetic
arm shows larger variations in pitch angle. In its inner part (east
of the galaxy centre) the field's pitch angle is
$-25\degr\pm7\degr$, while the outer part (south to west) the pitch
angle varies between $0\degr$ (azimuthal) and $-40\degr$ in a
quasi-periodical manner on a length scale of about 2\arcmin\
(3\kpc), where the structure function (Fig.~\ref{sf} top) starts to
rise.

Comparison of the pitch angles of the magnetic field vectors
(Fig.~\ref{pitch}) with the Faraday rotation measures (Fig.~\ref{rm}
left) reveals some degree of correlation. For example, $RM$ values
of $\le-100\radm$ in the south-east and at the western edge of the
map correspond to small absolute pitch angles in Fig.~\ref{pitch}.
Vice versa, $RM\ge100\radm$ about 3\arcmin\ west and east of the
centre correspond to large absolute pitch angles (about $-40\degr$).
Cross-correlation of the pitch angles of Fig.~\ref{pitch} with the
$RM$ data between \wave{3.5} and \wave{6.2} at the same resolution
yields a negative slope of $-$0.32 and a correlation coefficient of
$-$0.45. With more than 50 independent points and a Student-t
value of 4 this is a significant correlation. As $RM$ shows the
regular field component along the line of sight and the pitch angle
the field projection in the sky plane, systematic 3-D deviations of
the magnetic field lines from the plane of the galaxy are the
probable cause for the correlated deviations, such as \emph{helical
fields}\ (see Sect.~\ref{sect:depol}).

Although the field pitch angles are not constant in NGC~6946, they
may still be aligned with the spiral structures. To measure the {\em
pitch angles of the spiral structures}\, the wavelet technique of
Frick et al. (\cite{F00}) was applied to the polarized intensity map
(Fig.~\ref{pol} top right). The mean pitch angle of the northern
magnetic arm structure is $-25\degr\pm1\degr$ and that of the
southern magnetic arm is $-31\degr\pm2\degr$ (inner part, from east
to south) and $-15\degr\pm1\degr$ (outer part, from south to west).
Hence, the pitch angle of the magnetic field vectors and that of the
structure in polarized intensity agree well in the northern arm, but
only in parts of the southern arm. In M~51 the magnetic field in one
polarized arm has the same pitch angle (within a few degrees) as the
arm itself, but the field's pitch angle is systematically smaller by
10-20\degr\ than the pitch angle of the other polarized arm
(Patrikeev et al.\ \cite{P06}).

Alignment of the field vectors along the spiral structure can also
be achieved by compression of field and gas, e.g. in density waves.
However, the alignment of the field along the magnetic arms of
NGC~6946 cannot be explained by compression because any sign of gas
compression is missing. Density waves are relatively weak in
NGC~6946, probably due to the lack of a companion galaxy and of a
massive bar. Only in the inner region, at the inner edge of the
inner northern optical arm, density waves are sufficiently strong
for significant compression and field alignment along the optical
arm (Sect.~\ref{sect:compr}).

\begin{figure}[htbp]
\centerline{\includegraphics[bb = 62 128 528 623,width=0.495\textwidth,clip=]{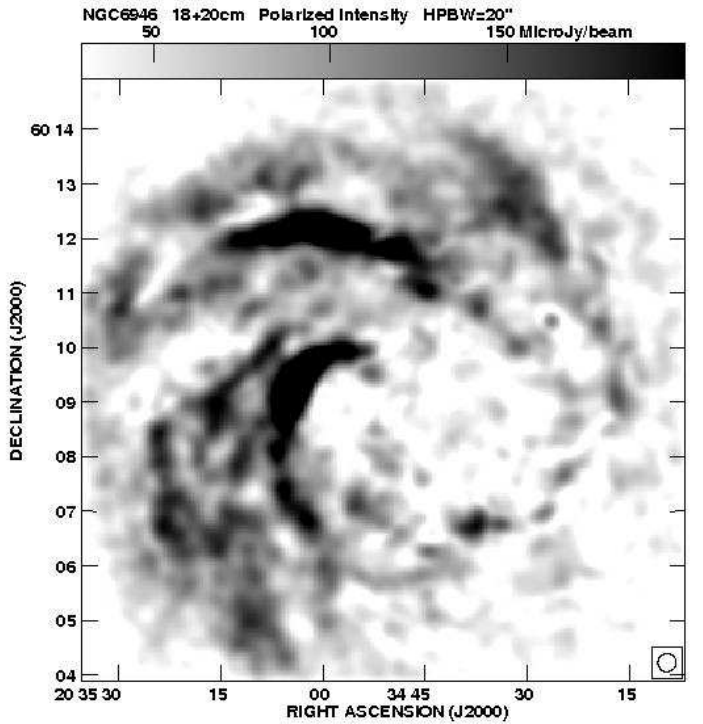}}
\caption{Average polarized intensity at \wave{18.0} and \wave{20.5}
at 20\arcsec\ resolution. } \label{pol20}
\end{figure}

The new observations presented in this paper raise challenges to
turbulent dynamo models. Firstly, two main magnetic arms in NGC~6946
are only the tip of the iceberg, the brightest parts of a network of
magnetic features (Figs.~\ref{HI} and \ref{pol20}). The
outer northern magnetic arms are also located between $\HI$ arms. A
dynamo origin of these outer arms is possible by the excitation of
azimuthal modes higher than $m=2$, but needs to be tested with
Faraday rotation measures, which however is not possible with the
present data. At wavelengths of \wave{18} and larger, the galaxy is
no longer transparent to polarized radio waves from the disk, so
that no reliable $RM$ can be determined. At shorter wavelengths, the
sensitivity of present-day radio telescopes is insufficient to
detect polarized synchrotron emission from the outer magnetic arms.
Polarized background sources of sufficiently small angular size do
not suffer from Faraday depolarization and can be used to determine
$RMs$ within the galaxy. With present-day telescopes the number
density of polarized sources is too low. Future measurements at long
wavelengths are favorable (e.g. with LOFAR) yielding a smaller $RM$
error.

Secondly, the total alignment of the field in the northern bright
magnetic arm is hard to understand because turbulent dynamos always
generate small-scale random fields simultaneously to large-scale
fields. Either the production of random fields is suppressed in the
magnetic arms, or these random fields are distributed smoothly over
the whole disk, observable as smooth, unpolarized emission. Future
turbulent dynamo models including the simultaneous generation of
small-scale and large-scale magnetic fields are required.

Thirdly, the partial alignment of the field along the spiral
structures and the network of magnetic arms call for nonlinear
dynamos. In linear dynamo models, the pitch angle of the magnetic
field vectors and that of the generated polarized spiral arm
structure are independent. In NGC~6946 and several other spiral
galaxies, these two pitch angles are mostly similar. Nonlinear
dynamo models are required where the field can back-react onto the
gas flow.

The comparison of energy densities in Sect.~\ref{sect:energies}
demonstrated that the magnetic fields control the flow of the
ionized gas and are of similar importance as the turbulent motions
of the neutral gas, possibly even dominating (Fig.~\ref{en}). An
influence of the magnetic field onto the gas spiral arms is well
possible.

A few steps steps have already been done along this way. Elstner et
al. (\cite{E00}) modelled the magnetic field in a spiral galaxy from
dynamo action and shear in a time-dependent gas flow, without
back-reaction onto the gas flow. The regular magnetic field was
shown to avoid the gas arms, forming magnetic arms, as observed, and
the spiral arm flow adjusts the field along the regions of high gas
density. However, the pitch angles of the magnetic field vectors are
too small in the interarm regions. Inclusion of the
magneto-rotational instability (MRI) as the main source of
turbulence in the outer parts of galaxies (Sellwood \& Balbus\
\cite{S99}; Dziourkevitch et al.\ \cite{D04}) provide relatively
strong seed fields which may enhance dynamo action and generate
mixed dynamo modes with magnetic arms. The model by Nishikori et al.
(\cite{N06}) includes MRI, but neglects spiral flows, dynamo action
and the input of turbulent energy by supernovae. Many magnetic arms
appear in their model, but with frequent reversals of the azimuthal
field direction along radius which are not observed. Furthermore,
the field also reverses from the disk to the halo, so that the $RM$
of polarized background sources should be small, which also is in
conflict with the observations presented in this paper.

Future MHD models should include the detailed gas flow (as obtained
from high-resolution $\HI$ observations) and allow for back-reaction
of the generated magnetic field onto the gas flow.

\subsection{The origin of the depolarization asymmetry}
\label{sect:depol}

The second unexpected discovery from radio polarization observations
of NGC~6946 was the large-scale asymmetry in Faraday depolarization.
The strongest depolarization occurs in the SW quadrant
(Fig.~\ref{dp}). An origin in the foreground of the Milky Way by
strong RM gradients can be excluded because the RM foreground is
smooth (Sect.~\ref{sect:rings}). Strong Faraday dispersion in the
Galactic foreground in a region of about 5\arcmin\ size is possible
but improbable. Faraday rotation at long wavelengths is also
enhanced in the SW quadrant (Beck\ \cite{B91}) which would cause
enhanced depolarization only if both phenomena are located in the
emitting volume within NGC~6946. Enhanced depolarization by enhanced
thermal density or field strength in the disk of NGC~6946 is
improble as the star formation rate is low in this part of the
galaxy. The remaining explanations are halo fields or field
geometry.

\begin{figure}[htbp]
\centerline{\includegraphics[bb = 33 207 521 637,width=0.495\textwidth,clip=]{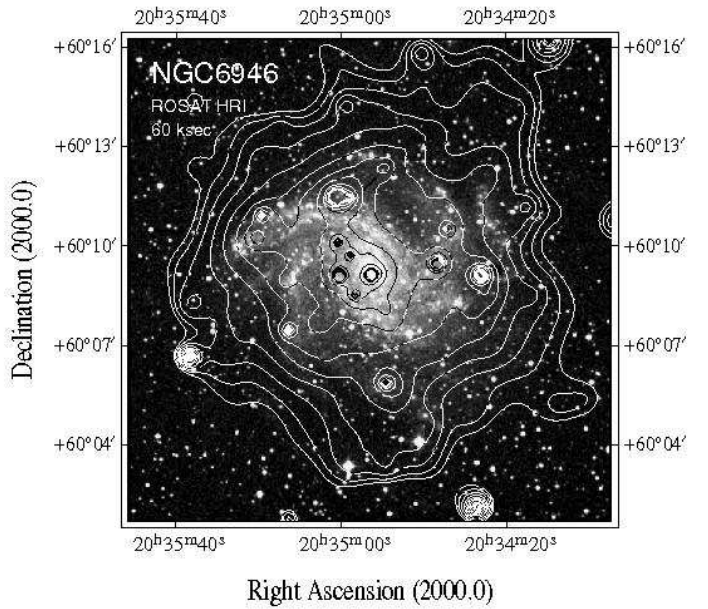}}
\caption{Deep 60~ksec ROSAT HRI image of NGC~6946 (Ehle, priv.
comm.) overlaid onto an optical image.} \label{xray}
\end{figure}

Faraday depolarization is generated by the field components along
the line of sight. The required difference between the vertical
field components on the eastern and western major axis cannot be
achieved by any symmetric field configuration. The asymmetry can be
interpreted as the effect of a vertical magnetic field (Beck\
\cite{B91}). This phenomenon may resemble \emph{solar coronal holes}
generated by fast streams of hot gas above regions of low
solar activity. In the centre of this region a huge stellar complex
with high wind speed is observed
(Efremov et al.\ \cite{E02}; Larsen et al.\ \cite{L02}).
If this is the origin of strong outflows of hot gas,
enhanced X-ray emission should be
observable from the halo in the SW quadrant of NGC~6946.
Figure~\ref{xray} shows a deep ROSAT image which reveals diffuse,
extended emission around the galaxy. The extension of the X-ray
emission is larger towards the south and towards the north-west,
but there is no significant increase in the region of strong
depolarization. More sensitive XMM data are forthcoming (Ehle \&
Beck, in prep).

A large-scale field asymmetry could also be caused by a
bisymmetric ($m=1$) dynamo mode which dominates in the halo.
However, Ehle \& Beck (\cite{EB93}) demonstrated that the rotation
measures between \wave{20.5} and \wave{2.8} (tracing disk plus halo)
can be fitted by a single-periodic sine wave of similar phase than
the RM between \wave{6.3} and \wave{2.8} (tracing the disk only),
excluding the presence of a bisymmetric mode with an amplitude
similar to those of the $m=0$ and $m=2$ modes.

\begin{figure}[htbp]
\centerline{\includegraphics[bb = 91 349 514 659,width=0.495\textwidth,clip=]{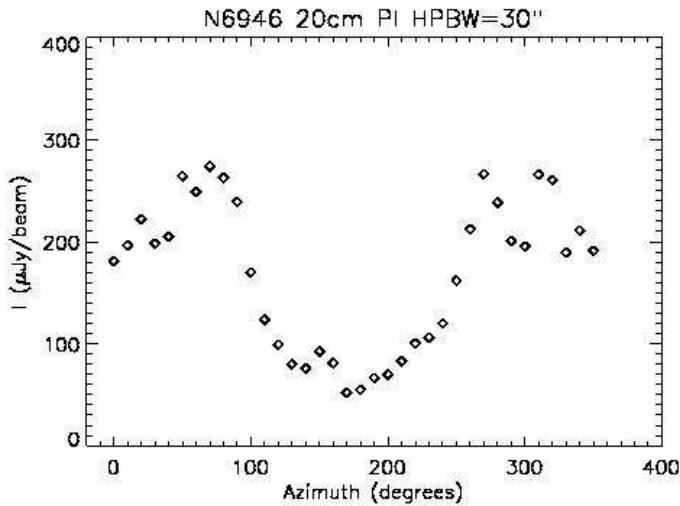}}
\caption{Polarized intensity at \wave{20.5} at 30\arcsec\
resolution, averaged in sectors of 10\degr\ width in azimuthal angle
in the galaxy plane and between 1\kpc\ and 10\kpc\ radius from the
centre of NGC~6946. The azimuthal angle is counted counterclockwise
from the galaxy's north-eastern major axis (at 60\degr\ position
angle). Errors due to rms noise are $\simeq5\muJyb$, smaller than
the symbol sizes.} \label{pi20}
\end{figure}

Interestingly, depolarization asymmetries exist in
several galaxies. The polarized intensity of NGC~6946 at \wave{20}
has a broad minimum centered on the south-western major axis (at
180\degr azimuth in Fig.~\ref{pi20}). The same phenomenon is
observed in M~83 which has a similar inclination and orientation in
space as NGC~6946 (Neininger et al.\ \cite{N93}). In IC~342 the
minimum at \wave{20} is also centered on the major axis, but on the
north-eastern side (Krause et al.\ \cite{K89}). Hence, all models of
asymmetric vertical fields are improbable because the field geometry
cannot depend on the direction to the observer. Taking into account
that the sense of winding of the spiral arms in IC~342 is reversed
compared to NGC~6946 and M~83, a common origin as an effect of the
{\em geometry of the field}\ with respect to the observer is
required. A possible explanation are \emph{helical fields}\
of sufficiently large diameter, so that Faraday depolarization of
emission from the far side of the helix is different from that from
the near side (Urbanik et al.\ \cite{U97}).
Detailed models are in preparation.

\section{Summary: NGC~6946, a gold mine of magnetic phenomena}

NGC~6946 is the ideal object to study magnetic fields with radio polarization.
Its size is large enough to allow high spatial resolution, and not too
large, so that it can be covered by a one single pointing with the VLA
at wavelengths larger than \wave{6}. NGC~6946 has one of the highest
star-formation rates among spiral galaxies and one of the strongest magnetic
fields. It is no surprise that this is the only spiral galaxy so far which has
been observed in radio polarization at five frequencies with high resolution
and high sensitivity.

The results on magnetic field properties in NGC~6946 and their
probable origins are summarized in Table~\ref{tab2}. The wealth of
different phenomena requires a combination of physical origins, like
dynamo action, density waves and shear flows. These phenomena occur
also in other galaxies, but with different weighting: Polarized
radio emission from M~31 is dominated by its coherent dynamo field
of axisymmetric type; the strong density waves of M~51 generate a
dominating anisotropic field by compression; the anisotropic fields
of the barred galaxies NGC~1097 and NGC~1365 are due to compression
and shear; the Large Magellanic clouds and the Magellanic-type
galaxy NGC~4449 hosts weak coherent dynamo fields, and small
irregular galaxies show turbulent fields only without signs of
coherency (see Beck\ \cite{B05} for a review). NGC~6946 offers
everything.

The low-frequency radio telescope LOFAR is under construction and
will map the magnetic fields in the outer disks of galaxies which
are populated by low-energy cosmic-ray electrons. The sensitivity of
the extended VLA (EVLA) will allow deeper mapping of galaxies in
polarization and Faraday rotation. Finally, the planned Square
Kilometre Array (SKA) will allow to search for magnetic fields in
distant galaxies and to accurately measure the spectrum of dynamo
modes in nearby galaxies (Beck\ \cite{B06}). With the SKA the number
density of polarized sources on sky will be increase by several
orders of magnitude and allow to measure a dense grid of $RMs$
through galaxies (Beck \& Gaensler\ \cite{BG04}). Thanks to its
large angular extent, NGC~6946 will be one of the prime targets for
these forthcoming radio telescopes.

\begin{table*}           
\caption{\label{Table2}Magnetic phenomena in NGC~6946 and their probable origins}
\centering
\begin{tabular}{ll}
\hline
Total magnetic energy density $\simeq$ turbulent energy density in inner galaxy & Field amplification by turbulent gas flows \\
Total magnetic energy density dominates in outer galaxy & Magneto-rotational instability ?\\
Low degrees of polarization in gas spiral arms & Field tangling due to star-formation activity \\
Large-scale pattern in Faraday rotation & Coherent fields \\
Magnetic arms with regular, almost completely aligned fields & Efficient dynamo action in interarm regions \\
Two inner magnetic arms & Superposition of dynamo modes $m=0$ and $m=2$ \\
No large-scale field reversals & Higher dynamo modes weak or absent in inner galaxy \\
Outer magnetic arms & Higher dynamo modes or shear flows in outer galaxy \\
Alignment of ordered fields along spiral structure & Interaction between gas flow and magnetic field \\
Magnetic arms are phase-shifted images of gas arms & Magnetic field may trigger the formation of gas spiral arms \\
North-south asymmetry in Faraday rotation & Spiral field points inwards in both inner magnetic arms \\
Enhanced magnetic field along inner edge of inner gas spiral arm & Compression or shear by density waves \\
Smooth turning of field lines into inner gas spiral arm & No shock \\
Mild field compression in inner gas spiral arm & Magnetic field connected to warm, diffuse gas \\
Anti-correlation of $RM$ and field pitch angle & Helical fields \\
$RM$ structure function flat until 6\kpc\ scale & Uncorrelated field structures below 6\kpc\ scale \\
$RM$ structure function rising for scales larger than 6\kpc & Large-scale coherent fields \\
Depolarization rings at \wave{20} & Differential Faraday rotation in $\HII$ complexes \\
Strong Faraday depolarization at \wave{18} and \wave{20} & Faraday dispersion by random fields \\
Large-scale asymmetry in Faraday depolarization & Helical fields ?\\
\hline
\label{tab2}
\end{tabular}
\end{table*}

\section*{Acknowledgements}
I wish to thank Elly M. Berkhuijsen, Andrew Fletcher and
especially Anvar Shukurov for useful discussions and many helpful comments.
I thank B{\"u}lent Uyan{\i}ker and Tom Pannuti for reducing the
Effelsberg
\wave{11} and \wave{21} observations, Matthias Ehle for providing
his ROSAT map, and B{\"u}lent Uyan{\i}ker for his software to compute
structure functions.


\end{document}